\documentclass[twocolumn,prb,aps,amsmath,amssymb,superscriptaddress,showpacs]{revtex4}

\usepackage{graphicx}
\usepackage{SIunits}
\usepackage{multirow}

\newcommand{\bfsigma}{{\boldsymbol{\sigma}}}
\newcommand{\bfepsilon}{{\boldsymbol{\epsilon}}}
\newcommand{\bfk}{\mathbf{k}}
\newcommand{\bfalpha}{\boldsymbol{\alpha}}
\newcommand{\bfR}{\mathbf{R}}
\newcommand{\bfE}{\mathbf{E}}
\newcommand{\bfB}{\mathbf{B}}
\newcommand{\bfp}{\mathbf{p}}
\newcommand{\bfr}{\mathbf{r}}
\newcommand{\bfc}{\mathbf{c}}

\newcommand{\bfA}{\mathbf{A}}
\newcommand{\bfS}{\mathbf{S}}
\newcommand{\bfL}{\mathbf{L}}
\newcommand{\calH}{{\cal{H}}}
\newcommand{\calD}{{\cal{D}}}
\newcommand{\calT}{{\cal{T}}}

\newcommand{\calO}{{\cal{O}}}
\newcommand{\calE}{{\cal{E}}}
\newcommand{\idroit}{\vert{i}\rangle}
\newcommand{\fgauche}{\langle{f}\vert}
\newcommand{\ftgauche}{\langle{\tilde{f}}\vert}

\begin{document}

\title{X-ray Magnetic and Natural Circular Dichroism from first principles:
Calculation of \textit{K}- and \textit{L}$_\mathit{1}$-edge spectra}

\author{N. Bouldi}
\affiliation{UPMC Univ Paris 06, CNRS, UMR 7590, IRD, MNHN, Institut de%
Min\'eralogie, de Physique des Mat\'eriaux et de Cosmochimie (IMPMC), %
case 115, 4 place Jussieu, 75252, Paris cedex 05, France}
\affiliation{Synchrotron SOLEIL, L'Orme des Merisiers, %
Saint-Aubin, BP 48, 91192 Gif-sur-Yvette Cedex, France}
\author{N. J. Vollmers}
\affiliation{Lehrstuhl f\"{u}r Theoretische Physik, Universit\"{a}t Paderborn,%
Warburger Str.~100, 33098 Paderborn, Germany}
\author{C.G. Delpy-Laplanche}
\affiliation{UPMC Univ Paris 06, CNRS, UMR 7590, IRD, MNHN, Institut de%
Min\'eralogie, de Physique des Mat\'eriaux et de Cosmochimie (IMPMC), %
case 115, 4 place Jussieu, 75252, Paris cedex 05, France}
\author{Y. Joly}
\affiliation{Institut N\'eel, CNRS and Universit\'e Joseph Fourier, BP. 166, 38042 Grenoble Cedex 9, France}
\author{A. Juhin}
\affiliation{UPMC Univ Paris 06, CNRS, UMR 7590, IRD, MNHN, Institut de%
Min\'eralogie, de Physique des Mat\'eriaux et de Cosmochimie (IMPMC), %
case 115, 4 place Jussieu, 75252, Paris cedex 05, France}
\author{Ph. Sainctavit}
\affiliation{UPMC Univ Paris 06, CNRS, UMR 7590, IRD, MNHN, Institut de%
Min\'eralogie, de Physique des Mat\'eriaux et de Cosmochimie (IMPMC), %
case 115, 4 place Jussieu, 75252, Paris cedex 05, France}
\author{Ch. Brouder}
\affiliation{UPMC Univ Paris 06, CNRS, UMR 7590, IRD, MNHN, Institut de%
Min\'eralogie, de Physique des Mat\'eriaux et de Cosmochimie (IMPMC), %
case 115, 4 place Jussieu, 75252, Paris cedex 05, France}
\author{M. Calandra}
\affiliation{UPMC Univ Paris 06, CNRS, UMR 7590, IRD, MNHN, Institut de%
Min\'eralogie, de Physique des Mat\'eriaux et de Cosmochimie (IMPMC), %
case 115, 4 place Jussieu, 75252, Paris cedex 05, France}
\author{L. Paulatto}
\affiliation{UPMC Univ Paris 06, CNRS, UMR 7590, IRD, MNHN, Institut de%
Min\'eralogie, de Physique des Mat\'eriaux et de Cosmochimie (IMPMC), %
case 115, 4 place Jussieu, 75252, Paris cedex 05, France}
\author{F. Mauri}
\affiliation{Departimento di Fisica, Universit\`a di Roma La Sapienza,%
 Piazzale Aldo Moro 5, I-00185 Roma, Italy}
\author{U. Gerstmann}
\affiliation{Lehrstuhl f\"{u}r Theoretische Physik, Universit\"{a}t Paderborn,%
Warburger Str.~100, 33098 Paderborn, Germany}

\date{\today}
\begin{abstract}
An efficient first principles approach to calculate X-ray 
magnetic circular dichroism (XMCD)
and X-ray natural circular dichroism (XNCD) is developed and applied in the near edge region at the \textit{K}- 
and \textit{L}$_\mathit{1}$-edges in solids. Computation of circular dichroism requires precise calculations of X-ray absorption spectra (XAS) for circularly polarized light.  For the derivation of the XAS cross section, we used a  
relativistic description of the photon-electron interaction
that results in an additional term in the cross-section 
that couples the electric dipole operator with an operator $\bfsigma\cdot (\bfepsilon \times \bfr)$ that we name
spin-position. The numerical method relies
on pseudopotentials, on the gauge including projected 
augmented wave method and on a collinear spin
relativistic description of the electronic structure.
 We apply the method to the calculations of
\textit{K}-edge XMCD spectra of ferromagnetic iron, cobalt
and nickel and of I  \textit{L}$_\mathit{1}$-edge XNCD spectra
of $\alpha$-LiIO$_3$, a compound with broken inversion symmetry. 
For XMCD spectra we find that, even if the electric dipole term 
is the dominant one, the electric 
quadrupole term is not negligible (8\% in amplitude in the case of iron). The term coupling the 
electric dipole operator
with the spin-position operator is significant (28\% in amplitude in the case of iron). We obtain a sum-rule 
relating this new term to the
spin magnetic moment of the \textit{p}-states.
In $\alpha$-LiIO$_3$ we recover the expected angular dependence of the XNCD spectra.  
\end{abstract}

\maketitle

\section{Introduction}

A dichroic (``two-coloured" in Greek) material has the property to absorb
light differently depending on its polarization. 
X-ray Circular Dichroism is the difference 
between X-ray absorption spectra (XAS) obtained from 
left and right circularly polarized light so it
describes the dependence of the absorption 
cross-section on the state of circularly-polarized light.

In a magnetic sample, the breaking of time-reversal symmetry  
permits X-ray Magnetic 
Circular Dichroism (XMCD). It is a powerful 
tool for the study of 
the magnetic structure of complex systems 
as it gives element-specific information.
Almost all synchrotron 
facilities around the world have a beamline 
dedicated to XMCD. \cite{rogalev_magnetic_2015}
The existence of well-established magneto-optical sum-rules 
that allow to obtain the spin and orbital
 contribution to the magnetic moment 
directly from the integral of the spectra  
\cite{thole_x-ray_1992,carra_x-ray_1993,altarelli_orbital-magnetization_1993}
 made it an essential technique to 
study the magnetic properties of matter. 
These sum-rules are widely and successfully applied at spin-orbit split 
\textit{L}$_\mathit{2,3}$-edges of transition metals
\cite{vogel_structure_1997,Stohr_exploring_1999,%
edmonds_ferromagnetic_2005,prado_magnetic_2013} 
and \textit{M}$_\mathit{4,5}$-edges of actinides. \cite{wilhelm_x-ray_2013}
On the other hand, in the absence of spin-orbit splitting of the core
state (as for \textit{K}- or \textit{L}$_\mathit{1}$-edge), 
only the orbital magnetization sum-rule 
\cite{thole_x-ray_1992,altarelli_orbital-magnetization_1993}
 can apply 
and a quantitative analysis of the spectra is far
from being straightforward.
Yet, for 3$d$ transition elements, measurements of 
XMCD at the \textit{K}-edge is the main way to 
probe magnetism under pressure and it is
a widely used technique despite the interpretation 
difficulties.
\cite{baudelet_x-ray_2005,torchio_pressure-induced_2011,cafun_xmcd_2013}

X-ray Natural Circular Dichroism (XNCD) occurs 
in non-centrosymmetric materials (for 
which the inversion symmetry is not a 
symmetry of the system). 
Up to now, it has been less widely used than XMCD but it 
presents a fundamental interest as it 
gives access to element specific stereochemical information.
\cite{rogalev_x-ray_2010}
In the domain of molecular magnetism, a renewed interest for this technique has recently
grown \cite{sessoli_strong_2015} with the emergence of new materials
that are both chiral and magnetic. 
Contrary to optical activity to which a large number
of mechanisms contribute,\cite{goulon_x-ray_1998} XNCD is
largely dominated by a single contribution.\cite{rogalev_x-ray_2010} 
At \textit{L}$_\mathit{1}$- and \textit{K}-edges, XNCD exists only if $p$ and $d$ orbitals are mixed,\cite{natoli_calculation_1998}
yielding a unique measure of the mixing
of even and odd orbitals.

The starting point of our work is a Density Functional Theory (DFT) based pseudopotential method. 
Using Projector Augmented Wave (PAW)
reconstruction, Lanczos algorithm and a continued fraction calculation, 
\cite{taillefumier_x-ray_2002,gougoussis_first-principles_2009,giannozzi_quantum_2009} it has proved successful for the 
calculation of absorption (XAS) spectra at the \textit{K}-edge. 
\cite{taillefumier_x-ray_2002,gougoussis_first-principles_2009,cabaret_first-principles_2010,bordage_electronic_2010} 
The \textit{L}$_\mathit{1}$-edge, that corresponds to a 
2$s$ core-hole, is expected to have the same behavior. 
In this paper, we propose the same kind of DFT-based approach
for the calculation of XMCD and XNCD spectra in the near-edge (XANES)
region.

Several calculations of XMCD at \textit{K}-edge in the XANES region can be found 
in the literature.
Most of these calculations are based on fully relativistic \cite{ebert_relativistic_1988, ebert_influence_1996,stahler_magnetic_1993,%
 gotsis_first-principles_1994, guo_what_1996%
, sipr_theoretical_2005} or semi-relativistic \cite{ brouder_multiple_1991,brouder_multiple_1996} 
multiple-scattering approaches with 
muffin-tin potentials  even if efforts have been made to go beyond this approximation.
 \cite{natoli_use_1986,joly_self_2009}

The technique presented in this paper allows the use of a free-shape
potential. Relativistic perturbations were taken into account both in the band structure\cite{gerstmann_rashba_2014} and in the photon matter interaction.\cite{bouldi_dirac}
The method has been implemented within a  highly efficient reciprocal space code that allows the 
modelling of a large range of systems.\cite{giannozzi_quantum_2009}

In section II, the terms that enter the absorption cross 
section up to the electric quadrupole approximation are listed. 
Section III is dedicated to the presentation 
of the computational method. Results obtained for XAS and XNCD at 
\textit{L}$_\mathit{1}$-edge  of iodine in both 
enantiomers of $\alpha$-LiIO$_3$ and for 
\textit{K}-edge XAS and XMCD spectra in 3$d$ ferromagnetic metals are
 presented in section IV. Finally, in section V, the 
relativistic operator is examined in 
detail within the collinear spin 
approximation.  Its corresponding sum-rule 
is derived and evaluated numerically and
an expression that allows 
for a simple implementation of this term is given.

\section{Contributions to the cross section}
\label{sec:contributions}
In the case of a fully circularly polarized light with a wave vector $\bfk$ along $z$,
the circular dichroism (XMCD and XNCD) cross section writes:
\begin{equation}
\sigma^{\mathrm{CD}}=\sigma(\bfepsilon_2)-\sigma(\bfepsilon_1)
\end{equation}
where $\bfepsilon_2=1/\sqrt{2}(1,i,0)$, $\bfepsilon_1=\bfepsilon_2^\star=1/
\sqrt{2}(1,-i,0)$ and $\sigma(\bfepsilon)$ is 
the X-ray absorption (XAS) cross section of the material.
XMCD effect at \textit{K}-edge of 3$d$ transition elements results at most 
in an asymmetry in absorption of the order of $10^{-3}$. 
For this study, it is therefore important to compute the absorption 
cross section very accurately.

In a monoelectronic semi-relativistic framework 
the contribution to the XAS cross section from a given
 core-state  of energy $E_i$ is given by (see appendix):
\begin{equation}
\sigma = 4\pi^2\alpha_0\hbar\omega
\sum_f |\langle f|T|i\rangle|^2
\delta(E_f-E_i-\hbar\omega)
\end{equation}
where $\alpha_0$ is the fine structure constant, $\vert  i\rangle$ is 
the two-components
wave function that corresponds to the large components of the Dirac 
wave function of the core state and the sum runs over unoccupied final
states  with energy $E_f$. The wave functions $\vert  f\rangle$ are eigenstates 
of the time-independent Foldy-Wouthuysen Hamiltonian of the electron
in the presence of an electromagnetic field 
$\bfE_0,\bfB_0$:\cite{itzykson_quantum_1980,bjorken_relativistic_1964}
\begin{multline}
H^{\mathrm{FW}} =   m c^2 +\frac{\bfp^2}{2m}
 + eV- \frac{e \hbar}{2m}\mathbf{\bfsigma}\cdot\bfB_0  \\
 -\frac{e\hbar}{4m^2c^2}\bfsigma \cdot 
   (\bfE_0\times \bfp )-\frac{e\hbar^2}{8m^2c^2}  \nabla\cdot\bfE_0 .
\end{multline}
Finally, $T$ is the sum of three operator: (i) the electric dipole operator, 
(ii) the electric quadrupole operator and (iii) a new light-matter interaction term that we named the spin-position operator:
\begin{equation}
T =\bfepsilon\cdot\bfr 
+ \frac{i}{2}
\bfepsilon\cdot\bfr ~ \bfk\cdot\bfr+ \frac{i\hbar\omega}{4mc^2}
\bfsigma \cdot (\bfepsilon \times \bfr)
\end{equation}
where $\bfsigma$ is the vector of Pauli matrices. 
 
The absorption cross section expands in six terms among which 
four terms are significant (see the orders of
magnitude in appendix).

The dominant term is the electric dipole-electric dipole (D-D) term:
\begin{equation}
\sigma_{\mathrm{D-D}}= 4 \pi^2 \alpha_0 \hbar \omega \sum_f \vert \langle f \vert
\bfepsilon \cdot \bfr \vert i\rangle \vert ^2 \delta(E_f-E_i-\hbar \omega).
\end{equation}
It is usually the only term that is taken into account in calculations of XAS 
and XMCD spectra at the \textit{L}$_\mathit{2,3}$-edges and sometimes at the \textit{K}-edge.
\cite{rehr_theoretical_2000,natoli_first-principles_1980,brouder_multiple_1996,fujikawa_relativistic_2003}

The electric quadrupole-electric quadrupole (Q-Q)  term is: 
\begin{multline}
\sigma_{\mathrm{Q-Q}}=  \pi^2 \alpha_0 \hbar  \omega \\ \sum_f \vert \langle f 
\vert (\bfk \cdot \bfr)(\bfepsilon \cdot \bfr) \vert i\rangle \vert ^2 
\delta(E_f-E_i-\hbar \omega).
\end{multline}
At the \textit{K}-edge, it can reach a few percent of $\sigma_{\mathrm{D-D}}$ 
in amplitude. It contributes mainly to the pre-edge region. It is sometimes included 
in X-ray absorption calculations. \cite{taillefumier_x-ray_2002,bunau_self-consistent_2009}

When neglecting spin-orbit coupling and in the absence of an external magnetic field, it is possible to choose real wave functions. 
In that case, the D-D and Q-Q terms verify
$\sigma(\bfepsilon)=\sigma(\bfepsilon^*)$, which leads to a 
zero contribution to circular dichroism. 
For this reason it is crucial to account for
relativistic effects in the wave functions calculation
in order to compute XMCD. 

On the other hand, the two following terms can give a non-vanishing
contribution to the circular dichroism cross section even when 
wave functions can be chosen real.

The electric dipole-electric quadrupole cross term (D-Q) is:
\begin{multline}
\sigma_{\mathrm{D-Q}}=-4\pi^2  \alpha_0\hbar\omega \\
\sum_f \Im[  \langle f \vert (\bfk \cdot \bfr)(\bfepsilon \cdot \bfr) 
\vert i\rangle\langle i \vert  \bfepsilon^{\star}\cdot \bfr  \vert f \rangle ]
\delta(E_f-E_i-\hbar \omega).
\end{multline}
If $\vert i \rangle$ and $\vert f \rangle$ are parity invariant (i.e. if 
inversion   $\bfr \rightarrow -\bfr$ is a symmetry of the system) 
then $\sigma_{\mathrm{D-Q}}=0$. 
It is however this term that is responsible for XNCD 
\cite{natoli_calculation_1998} because the electric dipole-magnetic dipole 
term (that is responsible for optical activity in the optical range) is very 
small in the X-ray range.

The cross term between the electric dipole and the relativistic operator that we named spin-position (D-SP) is:
\begin{multline}
\sigma_{\mathrm{D-SP}}=-\frac{2\pi^2 \alpha_0 \hbar^2 \omega^2  }{m c^2} \\
 \sum_f \Im[\langle f \vert \bfsigma \cdot 
 (\bfepsilon \times \bfr) \vert i\rangle
  \langle i \vert  \bfepsilon^{\star} \cdot \bfr  \vert f \rangle ] 
  \delta(E_f-E_i-\hbar \omega).
\label{dipSP}
\end{multline}
It exists only in magnetic materials. Like the spin-orbit coupling term in the 
FW Hamiltonian, it arises from the coupling of the small components of the 
Dirac wave functions.
To our knowledge, it has never been evaluated before.
We will show in the following that, despite the small prefactor of this term, 
its contribution to XMCD at the \textit{K}-edge of 3\textit{d} metals can account 
for up to one third of the 
XMCD intensity near the edge.

\section{Method \label{method}}

In the framework of the final state rule \cite{von_barth_dynamical_1982} the 
absorption cross section is obtained from one-electron wave functions. 
Within the frozen core approximation, the 1$s$ (\textit{K}-edge) or 2$s$ (\textit{L}$_\mathit{1}$-edge) 
unperturbed core states $\vert i \rangle$ can be determined from an all-electron isolated atom calculation. The stationary final states 
$\vert f \rangle$ are  calculated self-consistently in the presence of a core 
hole. Here, they are calculated within a semi-relativistic pseudopotential 
based DFT and Projector Augmented Wave (PAW) reconstruction framework.\cite{gerstmann_rashba_2014} The absorption cross-section is 
then calculated in a continued fraction scheme  using Lanczos algorithm.\cite{taillefumier_x-ray_2002,gougoussis_first-principles_2009} 

\subsection{Collinear semi-relativistic self-consistent field calculation}
Self-consistent field calculations in this study are based on Density Functional Theory 
(DFT) with a plane-wave basis set,  and pseudopotentials as implemented in 
\textsc{Quantum ESPRESSO} \cite{giannozzi_quantum_2009} including the 
spin-orbit coupling (SOC) term.\cite{gerstmann_rashba_2014} Since an accurate 
implementation of SOC plays a crucial role for the evaluation of XMCD 
spectra, we briefly describe the underlying approach in the following.

In pseudopotential-based methods the potential near the nuclei 
is replaced by a fictitious smooth potential.
The valence electrons wave functions are replaced by pseudo-wave functions 
that are exempt from the rapid oscillations near the core. The size of the 
plane-waves basis set needed to describe the system is therefore considerably 
lowered which leads to a much better computational efficiency compared to an 
all-electron approach making possible an \textit{ab initio} description of 
large systems with thousands of electrons.

In the PAW formalism, as described by Bl\"{o}chl,\cite{blochl_projector_1994}
the physical valence wave functions 
$\vert{\Psi}\rangle$ can be reconstructed from the pseudo-wave functions 
$\vert{\tilde{\Psi}}\rangle$ as they are related through a linear operator 
$\calT$: $\vert{\Psi}\rangle = \calT \vert{\tilde{\Psi}}\rangle$
with
\begin{equation}
\calT= 1\!\!1 + \sum_{\bfR,n} (\vert \phi_{\bfR,n}\rangle - 
\vert{\tilde{\phi}_{\bfR,n}}\rangle)\langle \tilde{p}_{\bfR,n}\vert.
\end{equation}

In our case, the set of all-electron partial waves centered on 
atomic site $\bfR$, $\vert \phi_{\bfR,n} \rangle$, are solutions 
of the Dirac equation for the isolated atom within a scalar relativistic 
approximation,\cite{koelling_technique_1977}
$\vert \tilde{\phi}_{\bfR,n} \rangle$ are the corresponding pseudo-partial waves 
and $\langle \tilde{p}_{\bfR,n}\vert$ form a complete set of projector 
functions. The operator $\cal{T}$ only acts in augmentation regions 
enclosing the atoms. Outside the augmentation regions the all-electron and 
pseudo-wave functions coincide. 

The pseudo-Hamiltonian is given by $\calT^\dagger H^{\mathrm{FW}} \calT$:\cite{ceresoli_first-principles_2010,gerstmann_rashba_2014}
\begin{equation}
\tilde{\calH}=E_{\text{kin}} + e \tilde{V}^{\text{loc}}(\bfr)+\sum_{\bfR} 
e\tilde{V}_\bfR^{\text{nl}}+\tilde{\calH}_{\text{SO}}
\end{equation}
where $E_{\text{kin}}$ is the kinetic energy as implemented in 
\textsc{Quantum ESPRESSO} and $\tilde{V}^{\text{loc}}$ and $\tilde{V}_\bfR^{\text{nl}}$ 
are the local part and the nonlocal part in separable form of the pseudopotentials. 
$\tilde{\calH}_{\text{SO}}$ is the pseudo-Hamiltonian corresponding to the 
time independent spin-orbit term in the Foldy-Wouthuysen transformed 
Hamiltonian:\cite{ceresoli_first-principles_2010}
\begin{align}
\tilde{\calH}_{\text{SO}}&= \calT^\dagger \left(  \frac{e\hbar}{4m^2c^2}
\bfsigma \cdot (\nabla V(\bfr){\times} \bfp ) \right) \calT \\ 
&= \frac{e\hbar}{4m^2c^2} \left(
 \bfsigma \cdot (\nabla \tilde{V}^{\text{loc}}(\bfr){\times}  \bfp) 
+ \sum_{\bfR} F_\bfR^{\text{nl}} \right). \nonumber
\end{align}
The $ F_\bfR^{\text{nl}}$ at the atomic site $\bfR$ are:
\cite{pickard_first-principles_2002}
\begin{multline}
F_\bfR^{\text{nl}}=\sum_{n,m}\vert \tilde{p}_n^\bfR \rangle \bfsigma \cdot
 (\langle \phi_{\bfR,n} \vert \nabla v_\bfR (\bfr){\times}  \bfp
  \vert \phi_{\bfR,n} \rangle \\ 
- \langle \tilde{\phi}_{\bfR,n} \vert \nabla \tilde{v}_\bfR^{\text{loc}}(\bfr) {\times} 
 \bfp \vert \tilde{\phi}_{\bfR,n} \rangle) \langle \tilde{p}_m^\bfR \vert 
\end{multline}
where $v_\bfR $ and $\tilde{v}_\bfR^{\text{loc}}$ are the atomic
all-electron and local channel pseudopotentials respectively. 
As these potentials are spherical, $ F_\bfR^{\text{nl}}$ rewrites:
\begin{multline}
F_\bfR^{\text{nl}}=\sum_{n,m}\vert \tilde{p}_n^\bfR \rangle \bfsigma \cdot
 (\langle \phi_{\bfR,n} \vert  \frac{1}{r}  \frac{\partial v_\bfR}{\partial r} \bfL
  \vert \phi_{\bfR,n} \rangle \\ 
- \langle \tilde{\phi}_{\bfR,n} \vert  \frac{1}{r}  \frac{\partial \tilde{v}_\bfR^{\text{loc}}}{\partial r} \bfL \vert \tilde{\phi}_{\bfR,n} \rangle) \langle \tilde{p}_m^\bfR \vert.
\end{multline}
The local potential $\tilde{V}^{\text{loc}}(\bfr)=\sum_{\bfR}\tilde{v}_\bfR^{\text{loc}}(\bfr)$
and the quantity $\frac{1}{r}  \frac{\partial \tilde{v}_\bfR^{\text{loc}}}{\partial r}$ decreases in $1/r^3$
so that the action of the operator $\tilde{V}^{\text{loc}}(\bfr){\times}  \bfp$ in
the augmentation region is, at first order, the same as the action of $\nabla \tilde{v}_\bfR^{\text{loc}}(\bfr){\times}  \bfp$.
In the PAW framework 
any pseudo-wave function in the augmentation region can be expanded according to 
$\vert \tilde{\Psi} \rangle= \sum_n \vert \tilde{\phi}_{n,\bfR} \rangle 
\langle \tilde{p}_n^\bfR  \vert \tilde{\Psi} \rangle$. 
Therefore, the term proportional to $\tilde{v}_\bfR^{\text{loc}}$ and the term 
proportional to $\tilde{V}^{\text{loc}}(\bfr)$ partially compensate each other
so that the dominant contribution arises from the term:
\begin{equation}
\frac{e\hbar}{4m^2c^2} \sum_{n \bfR m } \bfsigma \cdot
 \vert \tilde{p}_n^\bfR \rangle \langle \phi_{\bfR,n} \vert \frac{1}{r}  \frac{\partial v_\bfR}{\partial r}  \bfL \vert \phi_{\bfR,m} 
 \rangle \langle \tilde{p}_m^\bfR \vert.
\end{equation}
In this study, we consider collinear spin along $z$ and 
only the $z$ Pauli matrix is considered 
(diagonal spin-orbit coupling approximation):
\begin{equation}
\bfsigma= \sigma_z \mathbf{e}_z.
\end{equation}
In XMCD experiments a magnetic field is usually applied parallel to the beam,\cite{rogalev_x-ray_2006}
which justifies to consider the quantization axis parallel to $\bfk$.

This semi-relativistic 
approach, that includes spin-orbit coupling in a two-component approach, is 
computationally less expensive than a fully relativistic one.
It has been shown to reproduce the fully relativistic band structure.
\cite{gerstmann_rashba_2014}
For heavy atoms, the formula can be generalized by substituting $\nabla \tilde{V}^{\text{loc}}$
and  $\frac{\partial\tilde{v}_\bfR^{\text{loc}}}{\partial r}$
with reduced gradients, resulting in a ZORA-type 
of Hamiltonian.\cite{gerstmann_rashba_2014}

In this study, the calculations have been performed using Troullier-Martins 
norm-conserving pseudopotentials and are based on the generalized gradient 
approximation (GGA) with PBE functionals.\cite{perdew_generalized_1996} The charge 
density is evaluated self-consistently in the presence of a 
core hole which is described by removing a 
1\textit{s} or 2\textit{s} electron in the pseudopotential of the absorbing 
atom. A large unit cell (supercell) must be built to minimize the 
interactions between periodically reproduced core-holes and the \textit{k}-points grid 
can be reduced accordingly. 

\subsection{Cross-section calculation}

We implemented XMCD and XNCD in the \textsc{Xspectra} code\cite{gougoussis_first-principles_2009} of \textsc{Quantum ESPRESSO}\cite{giannozzi_quantum_2009} distribution.
The first results of this implementation for the terms D-D and Q-Q can be found in 
Ref.~\onlinecite{GougoussisPhD}.

In the PAW formalism it has been shown 
\cite{taillefumier_x-ray_2002,gougoussis_first-principles_2009} that the 
contribution of the operator $O$ to the absorption cross-section,
\begin{equation}
\sigma(\omega)=4 \pi^2 \alpha_0 \hbar \omega \sum_f 
\vert \fgauche O \idroit \vert ^2  \delta(E_f-E_i-\hbar \omega)
\label{crosssect_general}
\end{equation}
can be rewritten, as the initial wave function is localized around the 
absorbing atoms $\bfR_0$,
\begin{equation}
\sigma(\omega)=4 \pi^2 \alpha_0 \hbar \omega \sum_f 
\vert \ftgauche \tilde{\varphi}_{\bfR_0} \rangle \vert ^2  
\delta(E_f-E_i-\hbar \omega)
\label{crosssect_paw}
\end{equation}
with
\begin{equation}
 \vert \tilde{\varphi}_{\bfR_0} \rangle=\sum_{n} 
 \vert \tilde{p}_n^{\bfR_0} \rangle \langle \phi_n^{\bfR_0} 
 \vert O \idroit.
\end{equation}
This sum involves in principle an infinite number of projectors but 
experience demonstrated that two or three linearly independent projectors are in 
general sufficient in order to achieve the convergence of the D-D term at the 
\textit{K}-edge in the XANES region.\cite{bunau_projector_2013}  

The determination of all empty states in Eq.~\eqref{crosssect_paw} would 
require a lot of computing resources and, as a consequence, would limit the size of the 
manageable supercell. 
To increase the efficiency of the method, the cross section is evaluated as 
developed in Ref.~\onlinecite{taillefumier_x-ray_2002} and \onlinecite{gougoussis_first-principles_2009} via the Green's 
function using Lanczos algorithm\cite{lanczos_solution_1952}  which avoids 
the heavy workload of a large matrix inversion. 
The cross terms D-SP and D-Q are not in the form of Eq.~\eqref{crosssect_general}
but they can be determined from two calculations of this type 
using the relationship:
\begin{equation}
\Im[D B^\star] = \frac{1}{4} (|D+iB|^2-|D-iB|^2) 
\end{equation}
where $B$ is either the electric quadrupole or the spin-position operator and $D$ is the electric dipole operator.
For the term D-SP within the diagonal spin-orbit coupling approximation, we have checked that this approach yields the same result as 
the computational time sparing
calculation from the D-D spin-polarized contributions presented in section V 
(Eq. \eqref{linkDSP_DD}).

The calculated spectra are broadened with a Lorentzian function. 
Furthermore, the occupied states, that do 
not contribute to the absorption cross section, are cut according to the 
method described in paragraph III-B of 
Ref.~\onlinecite{brouder_multiple_1996}. 

For the selected examples below, the different contributions to the 
cross sections for left- and right-circularly polarized light $\sigma(\bfepsilon_2)$
and $\sigma(\bfepsilon_1)$ were 
computed accurately in order to obtain circular dichroism.

\section{Applications}

\subsection{Technical details}
\begin{figure}
\includegraphics[width=8cm]{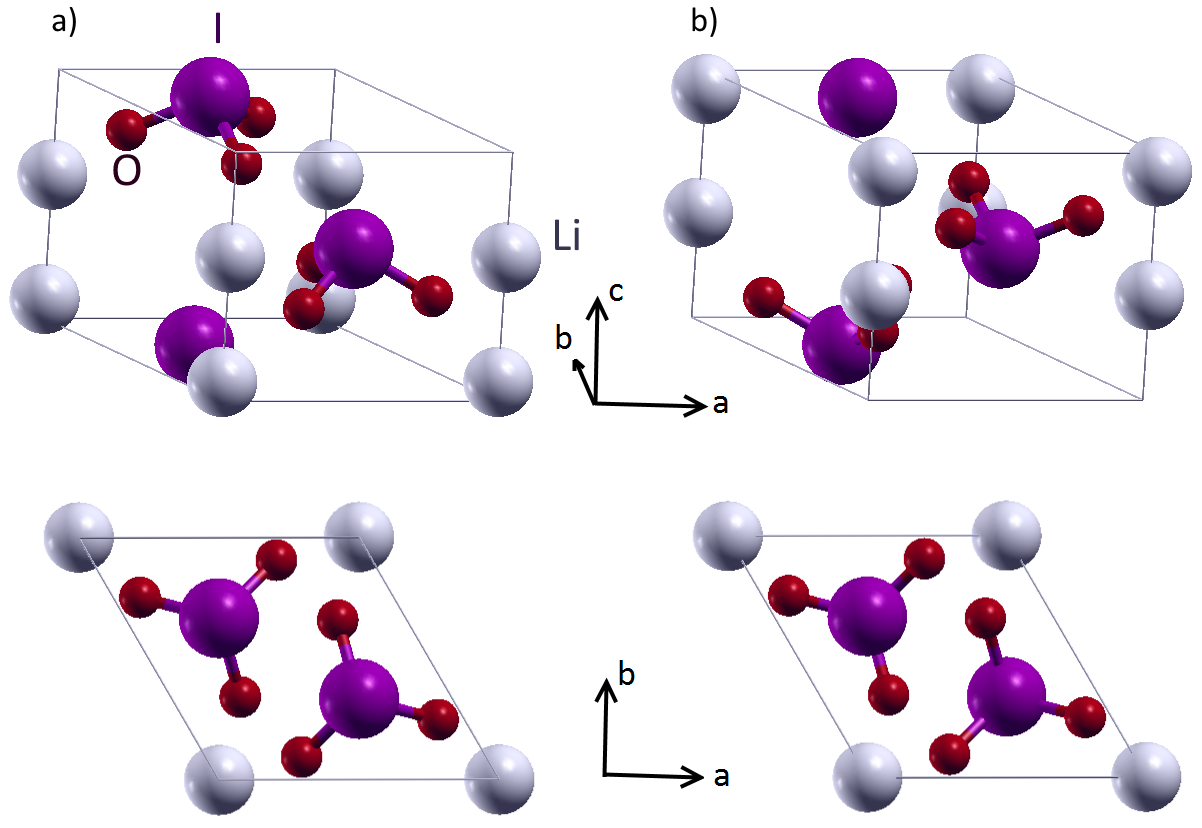}
\caption{Hexagonal $\alpha$-LiIO$_3$ unit cell for a) $\Delta$  and   b) $\Lambda$
 enantiomers.\cite{goulon_x-ray_1998}
On bottom: top view of the cells (projection on (001)).}
\label{LiIo3cells}
\end{figure}

For LiIO$_3$, the experimental structure is used:\cite{svensson_structural_1983} the
$\Delta$ enantiomer of
$\alpha$-LiIO$_3$ belongs to the hexagonal space group $P6_3$ with lattice 
parameter $a=5.48~\angstrom$ and $c=5.17~\angstrom$. The atomic positions 
\cite{goulon_x-ray_1998} are Li 2(a) (0,0,0.076), 
I 2(b) (1/3,2/3,0) and O 6(c) (0.247,0.342,0.838). 
The $\Lambda$ enantiomer is the mirror image of the 
$\Delta$ one (see Fig.~\ref{LiIo3cells}) and it belongs
to the same space group.
A $2 {\times} 2 {\times} 2$  supercell (80 atoms) is used so that the smallest distance 
between a core-hole and its periodic image is $10.344~\angstrom$. Gamma-centered \textit{k}-points grids
$3 {\times} 3 {\times} 3$  for the self-consistent charge density calculation 
and $9 {\times} 9 {\times} 9$ for the spectra calculation are used. A 
constant Lorentzian broadening, with full width at half maximum set to the core-hole lifetime broadening
3.46 eV,\cite{fuggle_unoccupied_1992} is applied. As XNCD is a structural effect and
not a magnetic effect, the calculation is not spin-polarized.

The XMCD calculations for the 3$d$ ferromagnetic metals are
  carried out by using the following experimental lattice parameters: $a=2.87~\angstrom$ for 
\textit{bcc} Fe, $a=3.52~\angstrom$ for \textit{fcc} Ni and $a=2.51~\angstrom$ 
and $c=4.07~\angstrom$ for \textit{hcp} Co. The number of atoms per supercell is 64 atoms
 for Fe and Ni  and 96 atoms for Co, so the smallest distance between the periodically repeated 
core-holes is $9.84~\angstrom$ in  Fe, $9.97~\angstrom$ in Ni 
and  $10.03~\angstrom$ in Co.
A Methfessel-Paxton cold smearing of 0.14 eV (0.01 Ry) and a centered $2 {\times} 2 {\times} 2$
\textit{k}-points grid are used for the self-consistent charge density calculation. 
The spectra calculation is performed with a  $6 {\times} 6 {\times} 6$ grid 
for Fe and Co and  a  $8 {\times}8 {\times} 8$ grid for Ni. These calculations are
performed with collinear spins along the easy axis of the crystals, that is to say, [001] for \textit{bcc} Fe and 
\textit{hcp} Co and [111] for \textit{fcc} Ni\cite{Ohandley_magnetism} and the wave vector $\bfk$ 
is set along the same axis.

The spectra are convolved with a Lorentzian broadening function to 
simulate the effect of the finite lifetime of the core-hole (constant in 
energy) and of the inelastic scattering of the photoelectron (additional energy-dependent broadening) for 
which we use the curves published by M\"{u}ller et \textit{al}.\cite{muller_x-ray_1982}

Experimental and calculated spectra are normalized such that the edge jump is equal to 1.

During the calculation of the spectra the origin of energy
is set to the Fermi energy of the material $E_F$.
For the spectra to be compared with experiment, a rigid shift in energy is applied to the 
calculated spectra to make the maxima of the calculated XAS correspond to the maxima of the experimental spectra. The same shift is applied to the XMCD spectra.
In the plots, the origin of energy $E_0$ is therefore the one chosen in the publications from which
the experimental spectra are extracted.

\subsection{XNCD at the \textit{L}$_\mathit{1}$-edge of I in $\alpha$-LiIO$_3$}

\begin{figure*}\centering
\includegraphics[width=10.8cm]{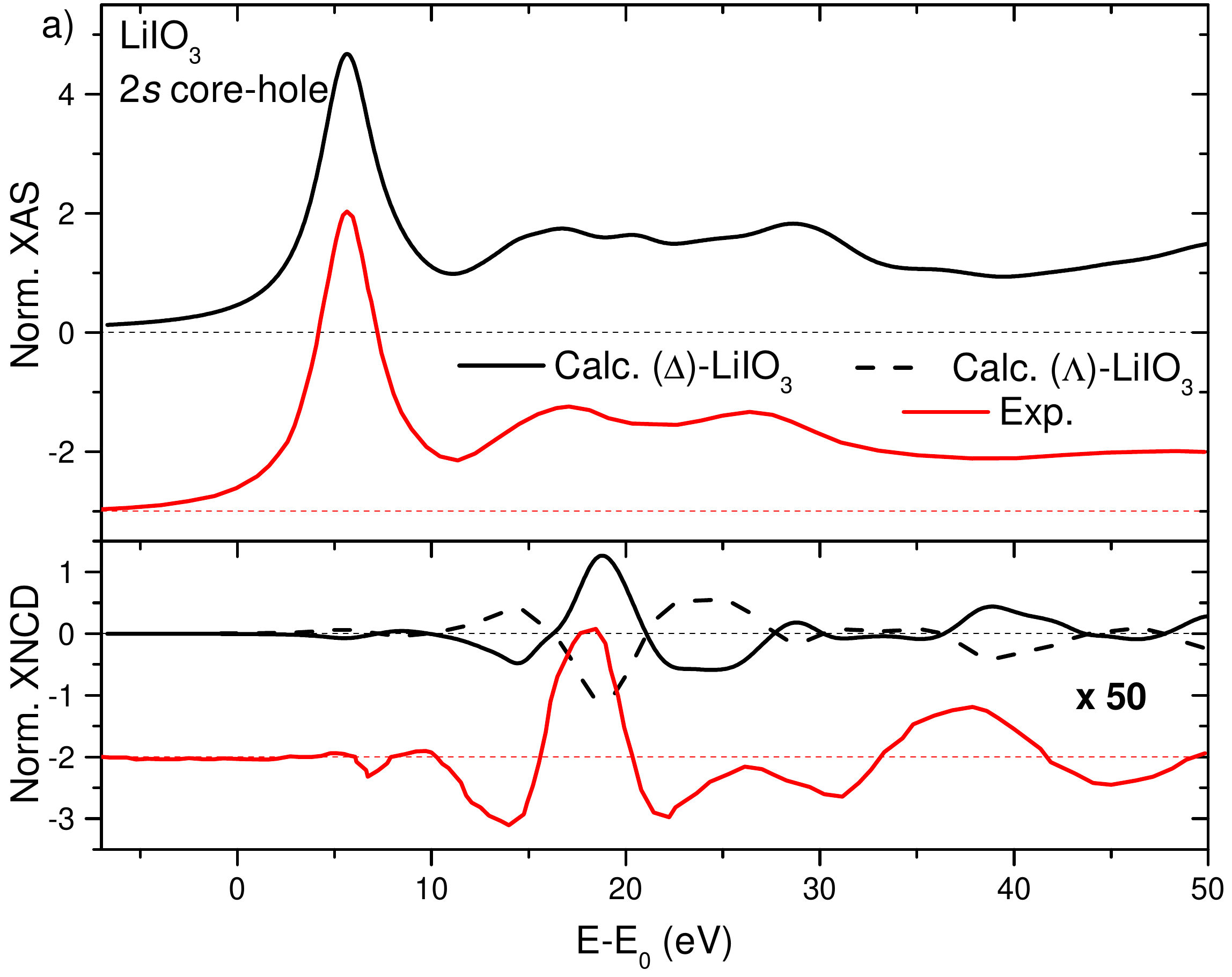}
\includegraphics[width=6.2cm]{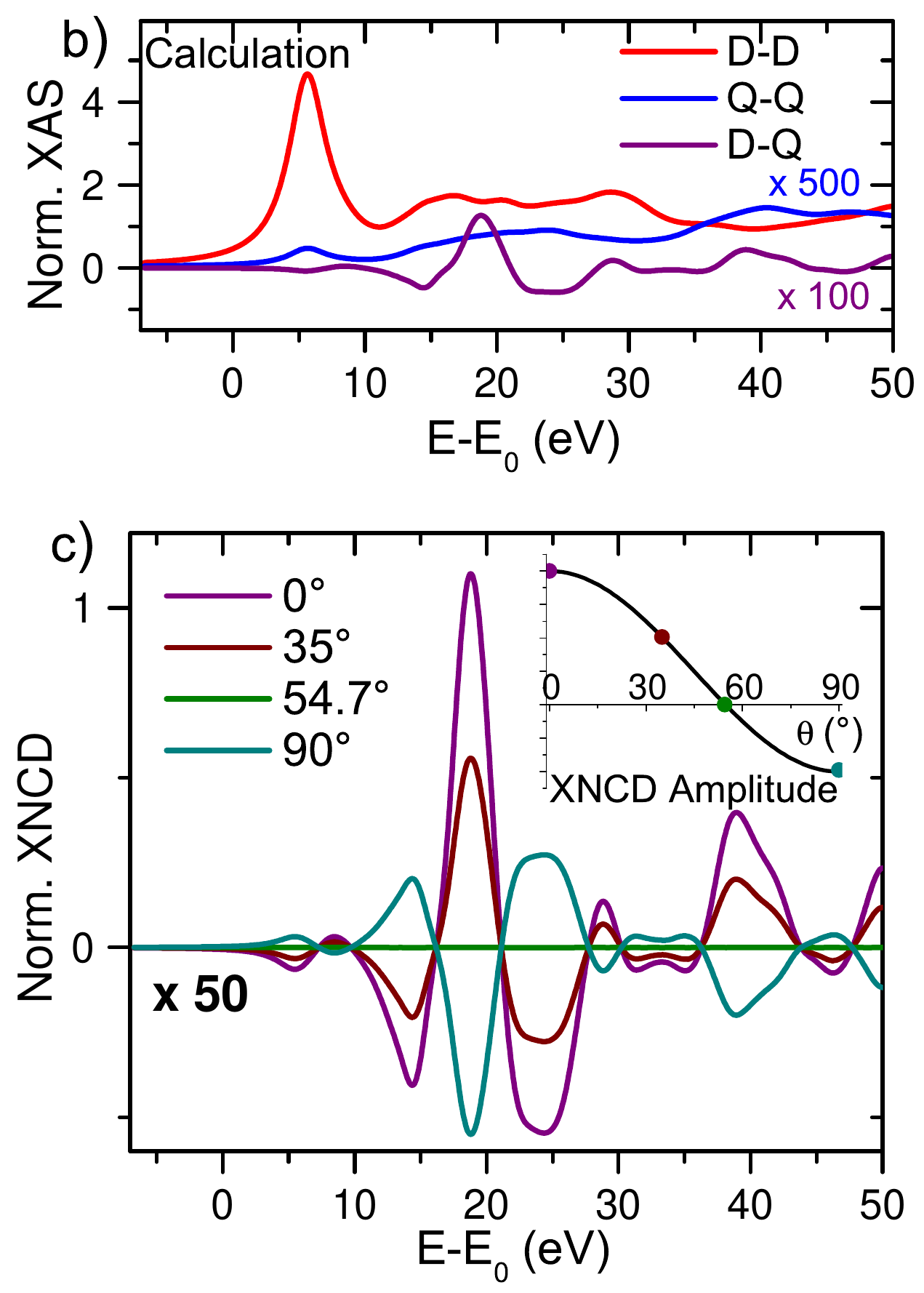}
\caption{a) Comparison of  experimental  \cite{goulon_x-ray_1998} and 
calculated XAS and XNCD spectra at the I \textit{L}$_\mathit{1}$-edge in LiIO$_3$ for both enantiomers 
 with $\bfk \parallel \bfc$. The XNCD spectra arises  from the D-Q 
term exclusively. Here in the calculation as in the experiment $\sigma^{CD}=\sigma^R-
\sigma^L$. 
b) Calculated contributions to the XAS at the I 
\textit{L}$_\mathit{1}$-edge in LiIO$_3$. The D-Q term was multiplied by 100 and the Q-Q term by 
500. 
c) Angular dependence of the XNCD at the I \textit{L}$_\mathit{1}$-edge in LiIO$_3$. In inset: 
XNCD amplitude as a function of the angle following the law 
$3\cos^2\theta-1$ where $\theta$ is the angle between $\mathbf{c}$ and the 
incident wave-vector $\mathbf{k}$.}
\label{LiIO3_I}
\end{figure*}

Natural circular dichroism in the inorganic non-centrosymmetric lithium 
iodate (LiIO$_3$) crystal have been measured in 1998 \cite{goulon_x-ray_1998} 
and it has been attributed to the interference of electric dipole and 
electric quadrupole transitions.\cite{goulon_x-ray_1998,natoli_calculation_1998} 
Previous calculations 
\cite{natoli_calculation_1998,goulon_x-ray_1998,ankudinov_theory_2000} 
were indeed able to reproduce the overall peak positions and intensities in 
this framework. The agreement is  however not entirely satisfactory for the absorption 
spectra. These discrepancies have been attributed to the use of muffin-tin 
potentials.\cite{ankudinov_theory_2000}

The approach presented in this work, that does not rely on the muffin-tin 
approximation, was applied to compute the XAS and XNCD spectra
for $\alpha$-LiIO$_3$.
The absorption is dominated by the electric dipole-electric dipole term (D-D)
as shown in Fig.\ref{LiIO3_I} b). 
The XNCD spectra, on the other hand, is entirely due to the 
cross term electric dipole-electric quadrupole (D-Q).

As illustrated by Fig.\ref{LiIO3_I} a), both the calculated XAS and XNCD spectra 
at the I \textit{L}$_\mathit{1}$-edge are in good agreement with experiment. 
However, the amplitude of the 
calculated XNCD is $4 {\times} 10^{-2}$ compared to the edge jump 
while the amplitude of the experimental spectra from  
Ref.~\onlinecite{goulon_x-ray_1998} is $6.5{\times} 10^{-2}$. Such an 
underestimation was also observed in Ref.~\onlinecite{natoli_calculation_1998}
within a multiple-scattering approach.

From Fig.\ref{LiIO3_I} a) bottom, it becomes obvious that 
the XNCD spectra for both enantiomers are opposite. Indeed, it has the same effect
for XNCD to change an enantiomer for the other ($\Delta \leftrightarrow \Lambda$)
as for XMCD to change the sign of the magnetic field ($\bfB \leftrightarrow -\bfB$). 

The angular dependence of the calculated XNCD spectra is depicted in 
 Fig. \ref{LiIO3_I} c) and its amplitude is plotted in inset as a function of 
$\theta$, the angle between $\bfk$ and the $\bfc$-axis of 
the crystal. 
This amplitude  varies as $3\cos^2\theta-1$ 
so it is maximal in the case $\bfk$ is parallel to the $\bfc$ axis.
This dependence is consistent with the formula derived in 
Ref.~\onlinecite{natoli_calculation_1998} for point group $C_6$ 
(point group of the space group of the crystal). 
Note that, as $\bfepsilon$ is kept perpendicular
to $\bfk$ and $C_6$ is a dichroic point group,\cite{brouder_angular_1990}
the XAS spectra also present an angular dependence. It does not prevent a
comparison of the amplitude of the XNCD spectra because the edge jump remains
unchanged. 

\subsection{XMCD at the \textit{K}-edge of 3d transition metals}

XMCD was recorded for the first time at the Fe \textit{K}-edge in magnetized Fe 
in 1987. \cite{schutz_absorption_1987} 
Ever since, a large number of calculations for the electric dipole term of 
the XMCD spectra on Fe \textit{K}-edge in \textit{bcc} Fe in the XANES region have been 
reported, for example in Ref.~\onlinecite{gotsis_first-principles_1994,%
igarashi_magnetic_1994,brouder_multiple_1996,guo_what_1996,%
ebert_influence_1996,fujikawa_relativistic_2003,sipr_theoretical_2005,dixit_ab_2016}. 
Calculations of XMCD at the \textit{K}-edge in \textit{fcc} Ni and \textit{hcp} Co are fewer
\cite{igarashi_magnetic_1994,igarashi_orbital_1996,guo_what_1996,torchio_x-ray_2011} 
and are not really conclusive.

These calculations have been performed with various methods, often within the 
electric dipole approximation and with muffin-tin potentials. Here, we present the calculation of the 
three terms (D-D, Q-Q and D-SP) that are likely to contribute to the XMCD cross-section at the \textit{K}-edge of 
 ferromagnetic 3\textit{d} transition metals showing the 
relevance of the D-SP term.

\begin{figure}[!h]
\includegraphics[width=8cm]{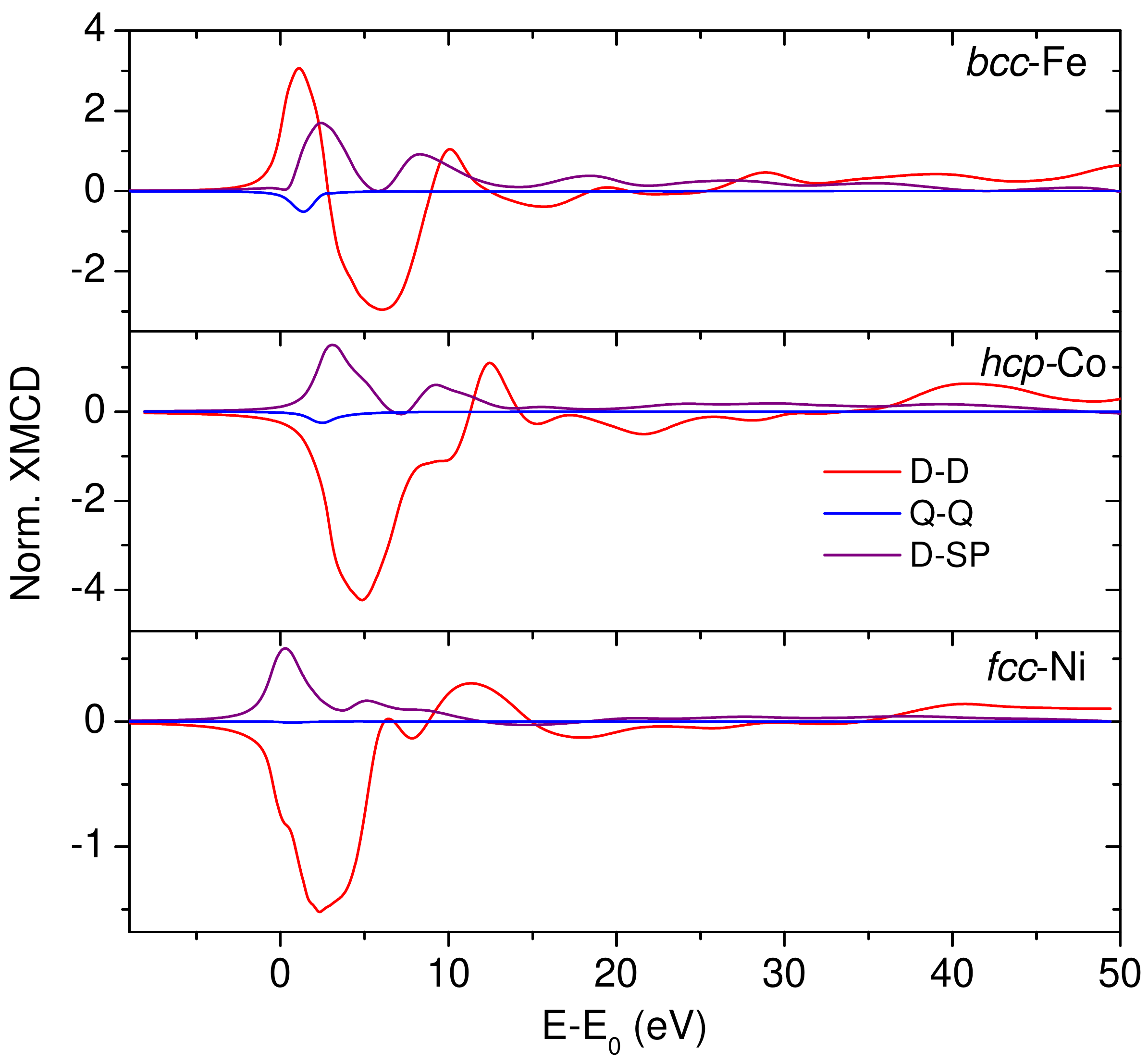}
\caption{Calculated contributions to the \textit{K}-edge XMCD spectra in the 
ferromagnetic 3\textit{d} metals Fe, Co and Ni.}\label{XMCD3dfe}
\end{figure}

The contribution of the D-SP term to the absorption cross-section is 
not shown here because it is
negligible. On the other hand its contribution to the XMCD 
spectra (Fig. \ref{XMCD3dfe}) is significant: it reaches 28\% of the D-D term 
in amplitude. This can be understood considering the sum-rules that are made 
explicit in the next section: in the XMCD cross-section, the D-SP term probes the spin 
polarization of the $p$ states whereas the D-D term probes their orbital 
polarization. In Ref.~\onlinecite{igarashi_orbital_1996} the $4p$ orbital 
magnetic moment in Co, Fe and Ni is
evaluated to a few $10^{-4}\mu_B$ 
(Fe: 5$\times 10^{-4}\mu_B$, Co: 16$\times 10^{-4}\mu_B$, Ni: 6$\times 10^{-4}\mu_B$) and in 
Ref.~\onlinecite{chen_experimental_1995} 
the  $4p$ spin magnetic moment in Fe and 
Co is evaluated to several $10^{-2}\mu_B$ 
(Fe: 5$ \times 10^{-2}\mu_B$, Co:6$ \times 10^{-2}\mu_B$) in the opposite direction.
This difference in order of magnitude of both quantities compensate for 
the smallness of prefactor ($\hbar \omega / 4 mc^2$) of the D-SP term (see
Table \ref{orderofmagnitude} in appendix).

\begin{figure}[!h]
\includegraphics[width=8cm]{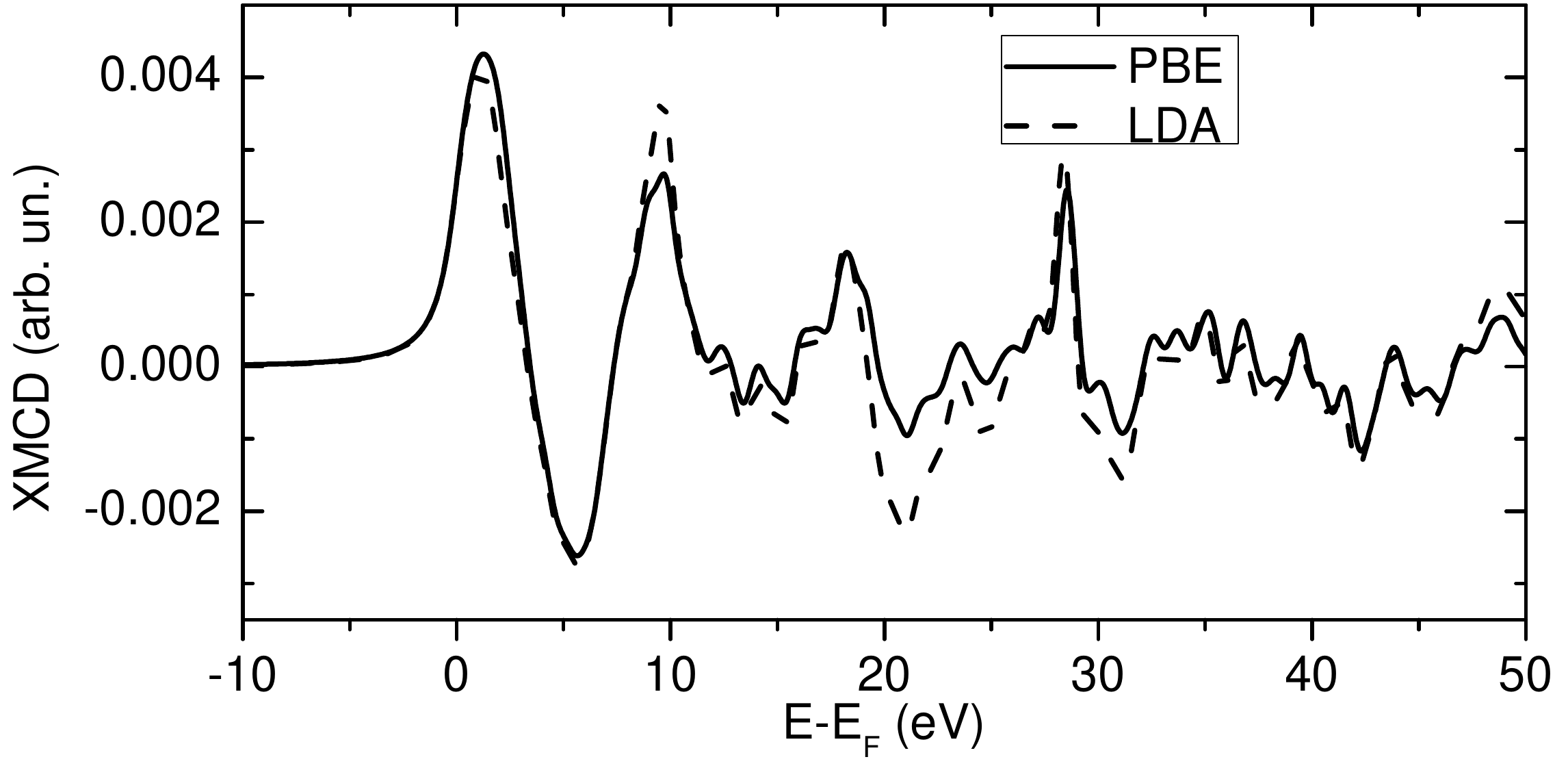}
\caption{Total calculated \textit{bcc} Fe \textit{K}-edge XMCD spectra without core 
hole using PBE and LDA functionals (all other technical parameters identical). Here, the broadening was taken constant (0.8 eV) along the whole energy range.}\label{ldapbe}
\end{figure}

\begin{figure*}
\includegraphics[width=8cm]{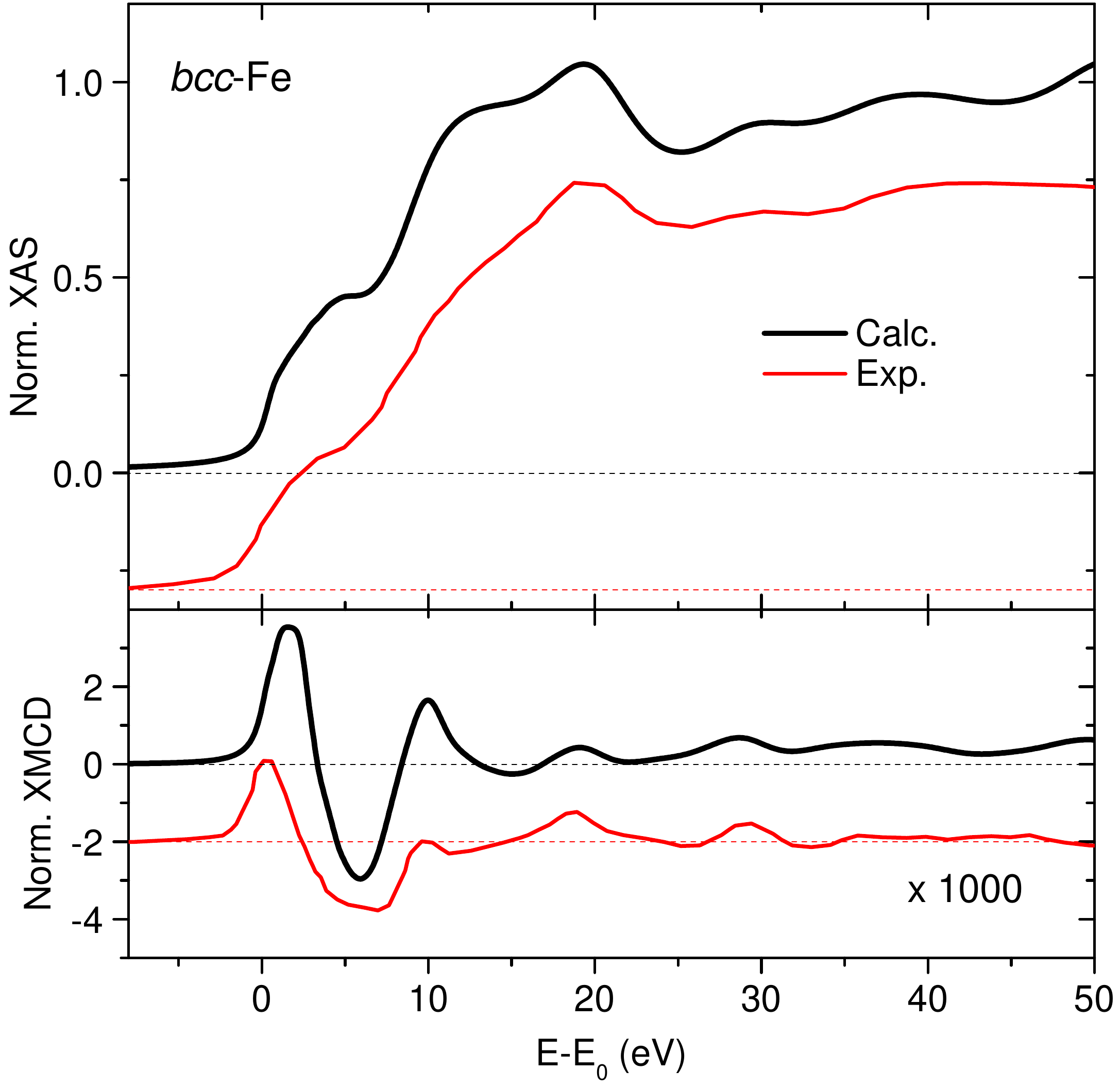}\hfill
\includegraphics[width=8cm]{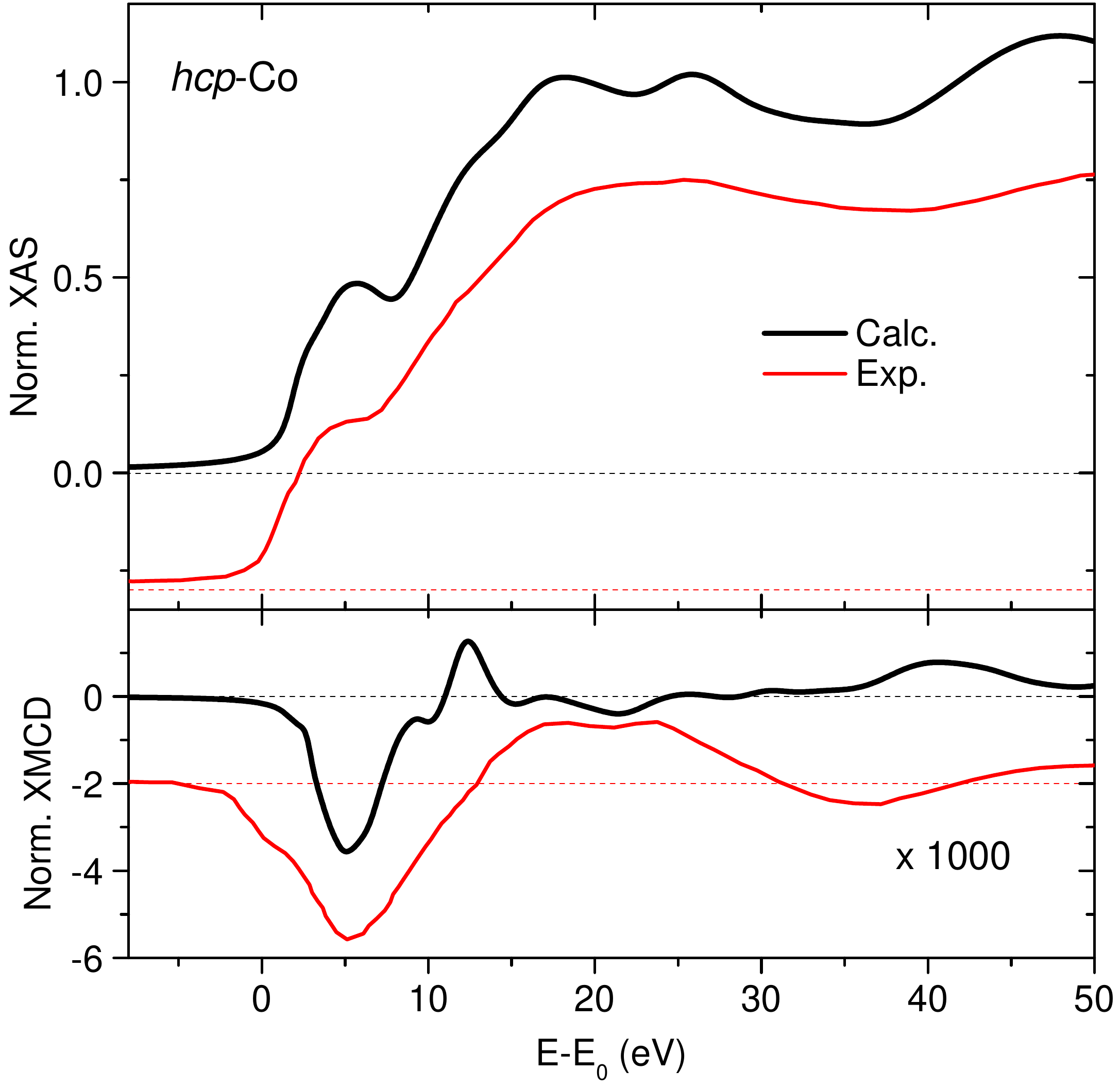}\hfill
\includegraphics[width=8cm]{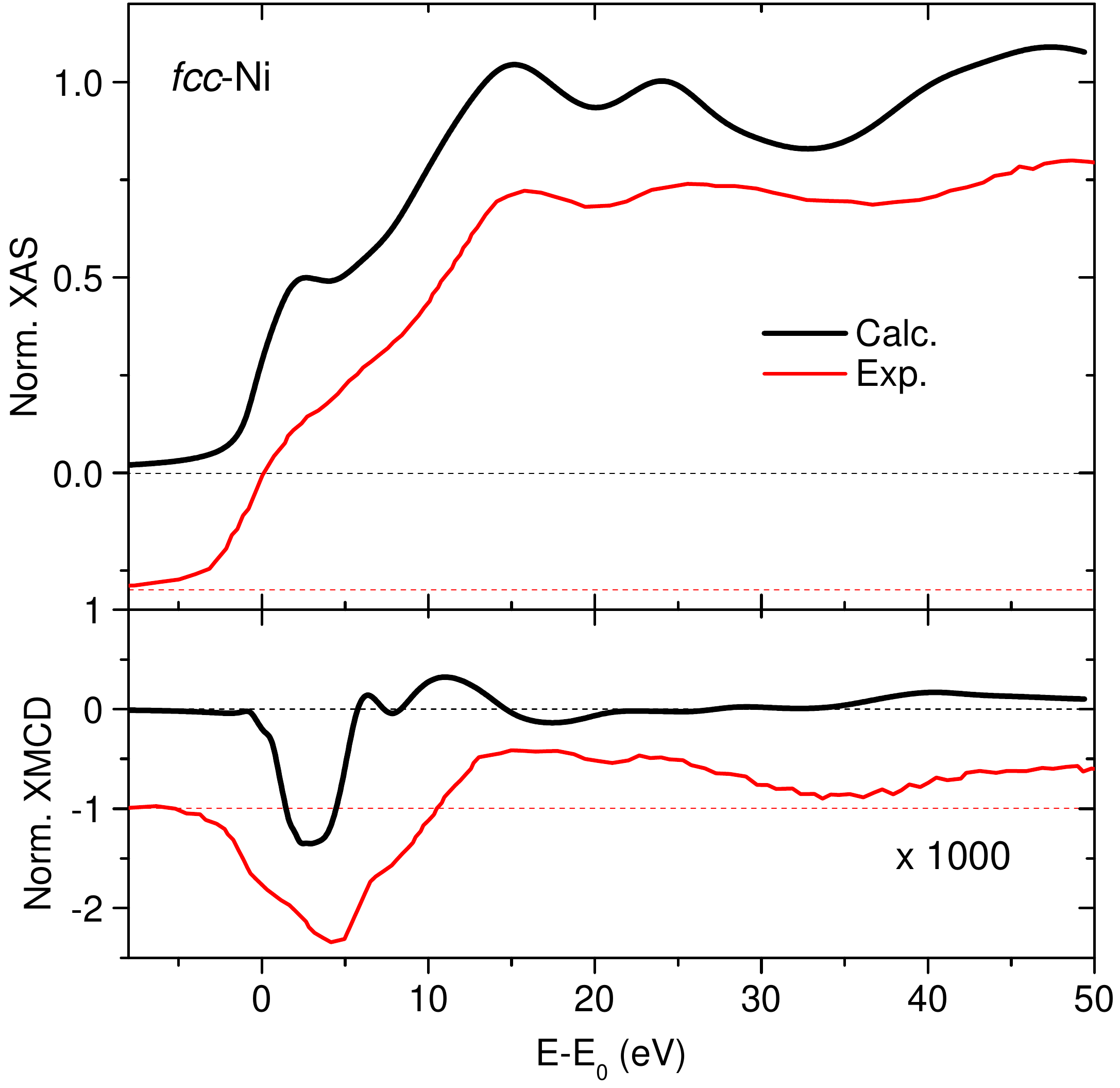}
\caption{Comparison between the experimental XAS and XMCD spectra for Fe, Co
\cite{laguna-marco_x-ray_2009,EXAFSdatabase} and Ni \cite{torchio_x-ray_2011} 
and the total calculated spectra. 
The wave vector and the magnetization axis were set to the easy axis 
of the crystals, that is to say, [001] for \textit{bcc} Fe and 
\textit{hcp} Co and [111] for \textit{fcc} Ni.\cite{Ohandley_magnetism}}\label{CalcExp3dfe}
\end{figure*}

To check possible numerical problems, we also performed the calculations using the FDMNES code\cite{bunau_self-consistent_2009} where, for this purpose, the D-SP term was introduced in the same way. This code follows Wood and Boring \cite{wood_improved_1978} to eliminate the small component and obtain
a couple of Schr\"odinger-like equations, including the spin-orbit effect, closely akin but improving the Pauli equation. Despite the very different approach (no pseudo-potential, calculation in real space and no diagonal spin-orbit coupling approximation), we found very similar results for both the shape and relative amplitude of the D-SP contribution. 

The agreement with the experimental spectra is fair as illustrated Fig. 
\ref{CalcExp3dfe}.  As usual in independent-particle calculations, the 
energy axis is slightly compressed 
\cite{materlik_ledge_1983,grunes_study_1983,lengeler_x-ray_1984} due to the 
energy dependence of the real part 
of the self-energy\cite{mustre_ab-initio_1991} 
for which corrections 
to the calculated spectra could be applied.\cite{kas_many-pole_2009}
Alternatively, the position of the calculated peaks
could be improved 
by phenomenological rescaling.\cite{mustre_ab-initio_1991,materlik_ledge_1983}

For Fe, the main peaks of the experimental XMCD 
are reproduced by the calculation. As in calculations by others,\cite{sipr_theoretical_2005} 
the positive peak at 10 eV is overestimated 
probably due to the approximate description of the
 exchange-correlation energy. Indeed, the comparison between the
  spectra calculated with PBE or LDA  functionals (Fig. \ref{ldapbe})
 shows that this peak would be even more enhanced with LDA.

For Ni and Co, a main negative peak is present near the main rising edge in the 
calculation as in the experiment but the satellite peaks that appear in the 
calculation are difficult to link to the experiment. 

In these calculations, the polarization rate of the light 
is taken to be 100 \% and a single crystal with full $3d$ spin polarization is considered. 
In Fe, Ni and Co, saturation 
is reached with usual experimentally applied magnetic field and the anisotropy
is quite weak so that the rate of circular polarization of the light $P_c$
is expected to account for most of the discrepancy in amplitude 
between the calculated and the experimental XMCD spectra. 
The data for Fe and Co were recorded in a 5 T magnet at 5 K and within a setup 
that reaches 90 \% circular 
polarization rate.\cite{BL39XUoutline} The correction on the amplitude of the
calculated spectra to fit the experimental condition should therefore be of 
order 0.9. Here, it is approximately 0.6 in the case of Fe and 1.0 in the case 
of Co. 
The data for Ni were recorded at ambient temperature in a 0.7 T magnet within 
a dispersive setup with a diamond quarter-wave plate for which we can infer 
that $P_c\approx 0.7$.\cite{Giles_energy_1994} However, no
correction on the amplitude of the calculated spectra is needed to make it
correspond to the amplitude of the experimental spectra. So, whereas our
calculation overestimates the amplitude of the XMCD spectra  in the case
of Fe, it underestimates it in the case of Ni.

\section{Contribution of the D-SP term to XMCD: the case of collinear spins}

\subsection{The SP operator}

In this section, we study  the \textit{spin-position} 
operator $SP(\bfepsilon)=\bfsigma \cdot (\bfepsilon \times \bfr)$.
We consider collinear spins along 
$z$ with independent spin channels. The spin part of the wave functions 
$\vert s \rangle$ can either be the spin up spinor $ \begin{pmatrix}
   1 \\
   0 
\end{pmatrix}$, or the spin down spinor $ \begin{pmatrix}
   0 \\
   1 
\end{pmatrix}$.

The D-SP term is the cross term between the electric dipole and the 
spin-position operator. Spin does not appear in the electric dipole operator, 
so it is diagonal in spin:
\begin{equation}
\langle \phi_i s \vert  \bfepsilon^{\star} \cdot \bfr  \vert \phi_f s' \rangle 
= \langle \phi_i \vert  \bfepsilon^{\star} \cdot \bfr  \vert \phi_f \rangle 
\delta_{s s'}.
\end{equation}
This imposes $s'=s$. 
On the  other hand, the vector of Pauli matrices $\bfsigma$ appears explicitly 
in the Spin-Position operator:
\begin{equation}
\langle \phi_i s \vert  \bfsigma \cdot (\bfepsilon \times \bfr)   \vert \phi_f 
s \rangle = \langle \phi_i \vert  (\bfepsilon \times \bfr)  \vert \phi_f 
\rangle  \cdot\langle s \vert \bfsigma  \vert s \rangle
\end{equation}
As $\langle s\vert \sigma_x  \vert s \rangle = 
\langle s \vert \sigma_y  \vert s \rangle=0$,
we can exclude a priori the terms that are proportional to $\sigma_x$ and 
$\sigma_y$ in the Spin-Position operator. 
In that case the spin position operator rewrites:
\begin{align}
SP_{\mathrm{col}}(\bfepsilon)&=\sigma_z 
(\epsilon_x y - \epsilon_y x)  \label{SPharm}\\
 &= \sigma_z\frac{4 i \pi}{3}r (Y_1^{-1}(\bfepsilon)Y_1^{1}(\mathbf{u_r})- 
 Y_1^{1}(\bfepsilon)Y_1^{-1}(\mathbf{u_r})). \nonumber
\end{align}
Its selection rules are almost the same as for the electric dipole 
\cite{sebilleau_x-ray_2006} one:
$\Delta l=\pm 1$, $\Delta m=\pm 1$.

As $Y_1^{-1}(\bfepsilon_1)=0$, $Y_1^{-1}(\bfepsilon_2)=\sqrt{3/4\pi}$,
 $Y_1^{0}(\bfepsilon_{1})=Y_1^{0}(\bfepsilon_{2})=0$, $Y_1^{1}(\bfepsilon_1)=-\sqrt{3/4\pi}$ and 
 $Y_1^{1}(\bfepsilon_2)=0$,
\begin{align}
SP_{\mathrm{col}}(\bfepsilon_1) &= i \sqrt{\frac{4\pi}{3}}r Y_1^{-1}
(\mathbf{u_r}) \sigma_z=\sigma_zi \bfepsilon_1. \bfr\\
SP_{\mathrm{col}}(\bfepsilon_2) &=  i \sqrt{\frac{4\pi}{3}}r Y_1^{1}
(\mathbf{u_r}) \sigma_z=- \sigma_z i \bfepsilon_2. \bfr.
\end{align}
Hence,
\begin{align}
\sigma_{\mathrm{D-SP}}(\bfepsilon_1)&=-\frac{\hbar \omega}{2 m c^2}
(\sigma^{\uparrow}_{\mathrm{D-D}}(\bfepsilon_1)-
\sigma^{\downarrow}_{\mathrm{D-D}}(\bfepsilon_1)) \\
\sigma_{\mathrm{D-SP}}(\bfepsilon_2)&=\frac{\hbar \omega}{2 m c^2}
(\sigma^{\uparrow}_{\mathrm{D-D}}(\bfepsilon_2)-
\sigma^{\downarrow}_{\mathrm{D-D}}(\bfepsilon_2))
\label{linkDSP_DD}
\end{align}
with
\begin{multline}
\sigma^{s}_{\mathrm{D-D}}(\bfepsilon)=  4 \pi^2 \alpha_0 \hbar  
\omega \\ \sum_f \vert \langle f^{s} \vert \bfepsilon \cdot 
\bfr \vert i^{s} \rangle \vert ^2 
\delta(E_f-E_i-\hbar \omega)
\end{multline}
where $s=\uparrow$ or $\downarrow$. 
Therefore, in the diagonal spin-orbit coupling collinear spins case, the D-SP term can be computed from the D-D 
cross section for the up and down spin channels. 

\subsection{Sum-rule at \textit{K}-edge}

A sum-rule is a formula in which the integral of the circular dichroism spectra due to a 
given term of the cross-section is expressed as a function of ground state 
expectation value of some operator.
The sum-rules at \textit{L}$_\mathit{2,3}$-edges are well established 
\cite{wu_first_1993,chen_experimental_1995} and are widely used 
to extract quantitative magnetic ground state properties. Their derivation is based on
several approximations among which the fact that the radial integrals are spin 
and energy independent.\cite{altarelli_sum-rules_1998}
At \textit{K}-edge the sum-rule for the electric dipole-electric dipole term 
\cite{ankudinov_sum_1995,carra_x-ray_1993,%
altarelli_orbital-magnetization_1993,igarashi_orbital_1996}
relates the integral of the XMCD spectra to the orbital magnetic moment of occupied $p$ 
states that is proportional to ${<}L_z{>}_p$. 
This sum rule is however almost impossible to apply in practice because the 
upper limit of the integral is not well defined and, in the case of 3$d$ 
transition elements, the 4$p$ states are almost unoccupied so ${<}L_z{>}_p$ is 
very small and has a minor impact on the magnetic moment of the material.
Deriving a similar sum-rule for the D-SP term is nevertheless very useful to 
understand why, despite its very small prefactor, this term is so large in 
XMCD. 
We derive it following the method of Thole et \textit{al.} 
\cite{thole_x-ray_1992,altarelli_orbital-magnetization_1993} with many body 
wave functions and operators assuming all spins collinear and within the diagonal spin-orbit coupling  
approximation.

In a many body framework, using the expression for $SP$ 
in terms of spherical harmonics Eq~\eqref{SPharm}:
\begin{multline}
\sigma_{\mathrm{D-SP}}(\bfepsilon)=\frac{2\pi^2\hbar^2\alpha_0\omega^2}{mc^2}
\sum_{\nu=-1}^{1} \Re [Y_1^{-\nu}(\bfepsilon^\star)\\
 \left( Y_1^{1}(\bfepsilon) \zeta_{\mathrm{D-SP}}^{1\nu}-Y_1^{-1}(\bfepsilon) 
 \zeta_{\mathrm{D-SP}}^{-1\nu}\right)]
 \label{DSPspher}
\end{multline}
with
\begin{multline}
\zeta_{\mathrm{D-SP}}^{\lambda\nu}=(-1)^\nu \left(\frac{4\pi}{3}
\right)^2\sum_f\langle f \vert \sum_i( {\sigma_z}_i r_i Y_1^{\lambda}
(\mathbf{u_{r_i}}))^\star  \vert g \rangle  \\
 \langle g \vert \sum_i r_i Y_1^{\nu}(\mathbf{u_{r_i}}) \vert f \rangle
 \delta(E_f-E_g-\hbar \omega).
\end{multline}

In a second quantized form with $l$,$m$ and $\sigma$ the usual quantum numbers:\cite{natoli_calculation_1998}
\begin{multline} 
\langle g \vert \sum_i r_iY_1^{\lambda}(\mathbf{u_{r_i}})\vert f \rangle =\sum_{l m \sigma l_0 m_0 \sigma_0'} 
 \sqrt{\frac{3(2l+1)}{4 \pi (2l_0+1)}} \\
(10l0\vert l_0 0)(1\lambda l m \vert l_0 m_0 ) \langle g \vert  a^\dagger _{l_0 m_0 \sigma}
a_{l m \sigma}  \vert f \rangle\calD_{l_0,l}
\end{multline} where
$\calD_{l_0,l}
 =   \int \mathrm{d} r \, r^3  R^{\star}_{l_0}(r) R_{l}(r)$ 
is assumed - as usual in sum-rules derivations - to be spin-independent.
The experimental procedure enables to obtain the signal corresponding to a 
specific $l_0$. At \textit{K}-edge $l_0=0$ and $m_0=0$ so that,
\begin{equation} 
 \langle g \vert \sum_i r_iY_1^{\nu}(\mathbf{u_{r_i}})\vert f \rangle=\sqrt{\frac{1}{4\pi}}\sum_{\sigma } 
 (-1)^\nu \langle g \vert  a^\dagger _{0 0 \sigma}a_{l {-\nu} \sigma}\vert f \rangle \calD
\end{equation}
where
$
\calD=\calD_{0,1}$.\\
Similarly, as  $\langle \sigma_0' \vert \sigma_z \vert \sigma'\rangle=\sigma' 
\delta_{\sigma_0',\sigma'}$
\begin{equation} 
 \langle g \vert \sum_i r_iY_1^{\lambda}(\mathbf{u_{r_i}})\sigma_{zi} \vert f \rangle=\sum_{\sigma'} 
\sigma' (-1)^\lambda \langle g \vert a^\dagger _{0 0 \sigma'}a_{l {-\lambda} \sigma'} \vert f \rangle \calD.
\end{equation}

Using the completeness relation $
\int \mathrm{d} E \, \sum_f |f\rangle\langle f |  \delta(E_f-E_g-E)= 1\!\!1 - 
|g\rangle\langle g|
$, as the core shell is full and under the assumption that the radial integral 
$\calD$ does not depend on energy:
\begin{multline}
\int \mathrm{d} E \, \zeta^{\lambda\nu}_{\mathrm{D-SP}} = \\ \frac{4\pi}{9} 
\sum_{\sigma}(-1)^\lambda \sigma \langle g \vert a_{1 {-\nu}\sigma}  
a^\dagger _{1 {-\lambda}\sigma} \vert g\rangle  |\calD|^2.
\label{SRzeta}
\end{multline}

The combination of Eq.~\eqref{SRzeta} and Eq.~\eqref{DSPspher} leads to:
\begin{multline}
\int \mathrm{d} \hbar \omega \,\frac{\sigma_{\mathrm{D-SP}}
(\bfepsilon_{\begin{subarray}{l}2 \\1 \end{subarray}})}{(\hbar \omega)^2} = \\ 
\frac{\pm 2\pi^2\alpha_0}{3mc^2}\vert \calD\vert^2  \langle g \vert a_{1 {\pm 1} 
\uparrow}  a^\dagger _{1 {\pm 1} \uparrow}-a_{1 {\pm 1} \downarrow}  
a^\dagger _{1 {\pm 1} \downarrow}\vert g\rangle  
\end{multline}
The difference between the two integrals yields the XMCD sum rule for the D-SP term:
\begin{equation}
\int \mathrm{d} \hbar \omega \,\frac{\sigma^{\mathrm{XMCD}}_{\mathrm{D-SP}}}
{(\hbar \omega)^2} =-\frac{2\pi^2\alpha_0}{3mc^2}\langle  Sz_{l=1}^{1,-1}  
\rangle\vert \calD \vert^2,
\label{DSP_sumrule}
\end{equation}
with the operator
\begin{equation}
Sz_{l=1}^{1,-1}=\sum_{m=-1,1}   a^\dagger_{1 m \downarrow}
a_{1 m \downarrow}  -  a^\dagger _{1 m \uparrow}a_{1 m \uparrow}
\end{equation}
corresponding to a partial 
spin polarization of the occupied $p$ 
states.

\begin{figure}[!h]
\includegraphics[width=8cm]{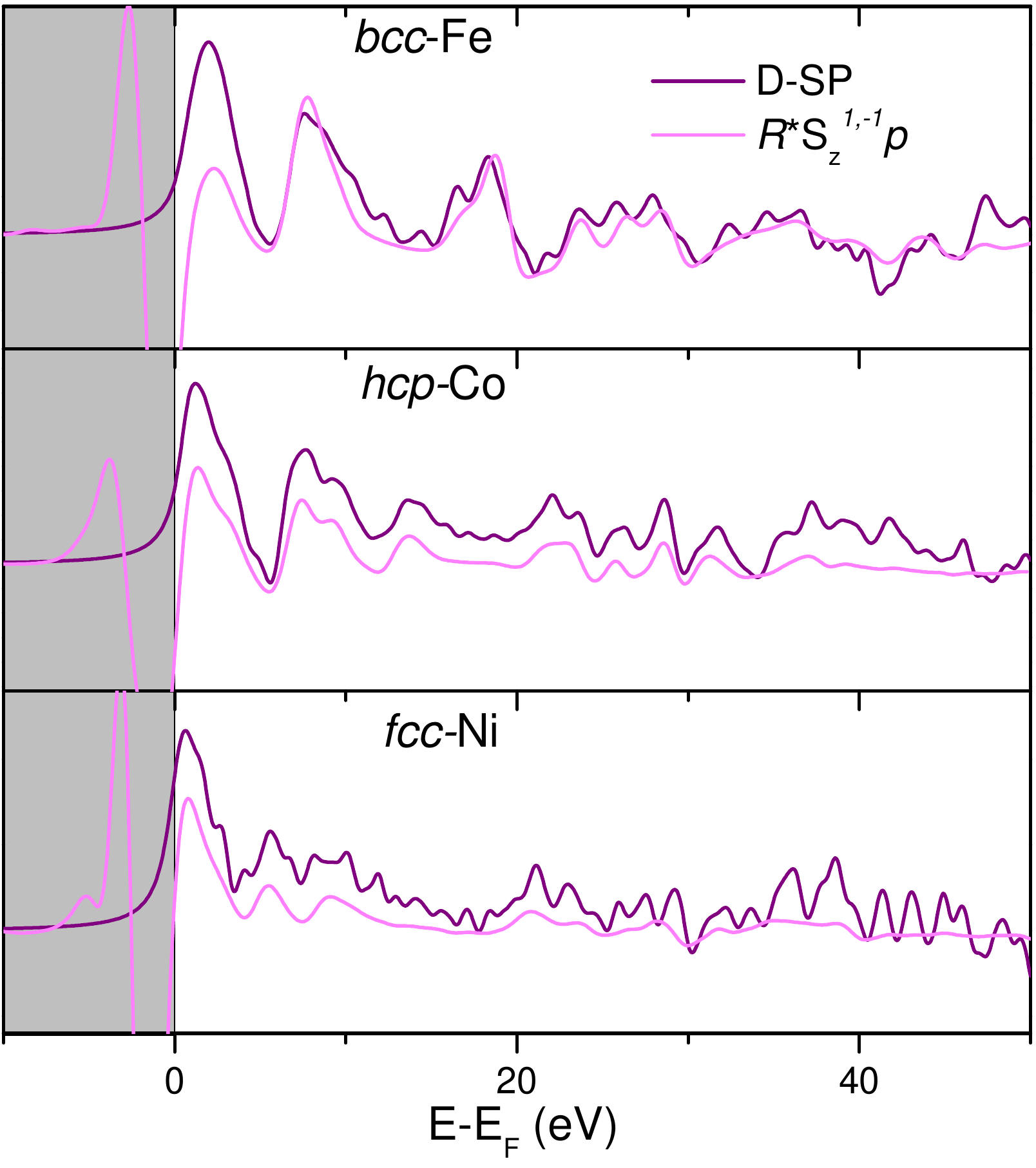}
\caption{Comparison between the calculated 
D-SP spectra without a core hole for Fe,Co 
and Ni and the calculated projected 
densities $Sz_{l=1}^{1,-1}(E)$. $Sz_{l=1}
^{1,-1}(E)$ has been multiplied by the 
factor between the $p$ density of states 
and the dipole XAS spectra times $R=
\frac{\hbar \omega}{2 mc^2}$ in accordance 
with the sum-rule Eq.~\eqref{DSP_sumrule}.}
\label{SumRules}
\end{figure}

If one considers the derivative of this 
sum-rule, we see that the electric dipole - spin position 
(D-SP) circular dichroism signal probes the spin 
polarization of the empty $p$ states. 
Fig. \ref{SumRules} illustrates the
correspondence between both quantities. This 
proves the validity of the D-SP sum-rule. 
Unfortunately, this sum rule can not be applied
directly on experimental spectra, mainly because of the 
superposition of the D-D contribution to the D-SP contribution. 

\section{Conclusion}

We have developed an efficient computational approach to determine accurate XMCD and XNCD 
spectra.

The main result is that the contribution from the relativistic term D-SP in the
 transition operator is significant in XMCD 
 spectra despite being negligible in XAS.
This importance is explained by the fact that 
this term probes the spin of the $p$ 
states that is two orders of magnitude larger than its orbital counterpart.  

For XNCD,
the calculated spectra  are in 
good agreement with experiment and the angular 
dependence corresponds to the expected one.

A big advantage of the method employed in this paper to perform XMCD and XNCD calculations 
is its wide adaptability that opens 
opportunities for applications to several 
kinds of systems such as strongly-correlated materials or molecules absorbed on functionalized surfaces. 
The same method could be apply to compute X-ray magneto-chiral dichroism (XM$\chi$D)
that has been observed in magnetized chiral systems \cite{sessoli_strong_2015}.
The features of XM$\chi$D differ from the one of XMCD and XNCD making
it a promising probe of the interplay between chirality and magnetism. 

\section*{Acknowledgement}
This work was supported by French state funds managed by the ANR within the 
Investissements d'Avenir programme under Reference No. ANR-11-IDEX-0004-02, 
and more specifically within the framework of the Cluster of Excellence MATISSE 
led by Sorbonne Universit\'es.
We are grateful to Delphine Cabaret for very interesting 
and constructive feedback on this work. 
We also thank Fran\c cois Baudelet and Lucie Nataf for providing reference spectra for the energy scaling.
UG and NJV acknowledge support by DFG (FOR 1405). The numerical calculations have 
been performed using HPC resources from the Paderborn Center for Parallel Computing (PC$^2$) and from GENCI-IDRIS
(Grant i2016096863).

\appendix*

\section{Semi-relativistic transformation of the relativistic cross section}

We start from the expression for the 
cross section in a relativistic framework \cite{bouldi_dirac} and 
we adapt it to the specific need of our numerical method that is 
the determination of large components of the Dirac wave function for the core state 
and of Foldy-Wouthuysen (FW) wave functions for the valence states.

\subsection{Relativistic cross section}
The contribution to the X-ray absorption (XAS) cross section from a given
 four-components Dirac core-state $\vert  
\Psi_i\rangle$ of energy $E_i$ is given by:\cite{bouldi_dirac} 
\begin{equation}
\sigma(\hbar \omega)=4 \pi^2 \alpha_0 \hbar  \omega \sum_f 
|\langle  \Psi_f \vert   T_D \vert  
\Psi_i\rangle \vert ^2 \delta(E_f-E_i-\hbar \omega)
\end{equation}
 where  the sum runs over unoccupied final states $\vert  
\Psi_f\rangle$ with energy $E_f$, $\alpha_0$ is the fine structure constant
 and $T_D$ is the transition operator defined as:
\begin{equation}
T_D =
\bfepsilon\cdot\bfr 
+ \frac{i}{2}
\bfepsilon\cdot\bfr ~ \bfk\cdot\bfr 
-
\frac{\hbar c}{2\omega} (\bfepsilon\times\bfk)\cdot(\bfr
\times\bfalpha)
\end{equation}
where the polarization vector $\bfepsilon$, the wave vector $\bfk$
and the energy $\hbar \omega$ describe the incident 
electromagnetic wave, $\bfr$ is the position operator 
and $\bfalpha=(\alpha_x,\alpha_y,\alpha_z)$ is
the vector of Dirac matrices. 

Here, as in our numerical calculations, a one-electron scheme is used.
In a many-body framework the formula for the cross-section would be the same 
but with N-electrons wave functions and many-body operators that write as sums 
over electrons. 

In Ref.~\onlinecite{bouldi_dirac}, the transformation into a two-component 
representation for $\vert \Psi_i\rangle$ and $\vert  
\Psi_f\rangle$ was performed by applying a time-independent
Foldy-Wouthuysen transformation (FW) at order $c^{-2}$. 
The FW transformation of $ \Psi_l$ is obtained by applying a
unitary operator: $ \psi_l^{\mathrm{FW}}=U_{\mathrm{FW}}  \Psi_l$ with,\cite{Eriksen_Foldy-Wouthuysen_1958}
\begin{equation}
U_{\mathrm{FW}} =1+\frac{\beta}{2mc^2}\calO-\frac{1}{8m^2c^4}\calO^2
\label{Ueriksen}
\end{equation}
where  $\beta$ the standard Dirac matrix. In this expression, $\calO$ is the odd operator
entering the Dirac Hamiltonian: $\calH^D=\beta mc^2+\calO+\calE$ where $\calE$ is even.
It is defined as
$
\calO=c\bfalpha\cdot(\bfp-e\bfA_0)
$
where $\bfp$ is the momentum operator
 and  $\bfA_0$ is the static vector potential.

Only the large components of $\psi_l^{\mathrm{FW}}$, denoted $\phi_l^{\mathrm{FW}}$, 
are non zero  up to order $c^{-2}$.
The cross-section can be written as a function of the large components
of $\psi_i^{\mathrm{FW}}$ and $\psi_f^{\mathrm{FW}}$:\cite{bouldi_dirac}
\begin{align}
\sigma &= 4\pi^2\alpha_0\hbar\omega
\sum_f |\langle \phi_f^{\mathrm{FW}}|T_{\mathrm{FW}}|
\phi_i^{\mathrm{FW}}\rangle|^2
\delta(E_f-E_i-\hbar\omega)
\label{CSFW}
\end{align}
The operator $T_{\mathrm{FW}}$ is the projection on upper components
of $U_{\mathrm{FW}} T_D U_{\mathrm{FW}}^\dagger$:
\begin{equation}
T_{\mathrm{FW}} = T_{\mathrm{D}}+T_{\mathrm{Q}}+T_{\mathrm{MD}}+T_{{a_0}}+T_{\mathrm{SP}}
\label{TFW}
\end{equation}
where
\begin{equation}
T_{\mathrm{D}}=\bfepsilon\cdot\bfr 
\end{equation}
and
\begin{equation}
T_{\mathrm{Q}}=\frac{i}{2}
\bfepsilon\cdot\bfr  ~  \bfk\cdot\bfr
\label{TQ}
\end{equation}
are the standard \textit{electric dipole} and 
\textit{electric quadrupole} operators.

The \textit{magnetic dipole} operator $T_{\mathrm{MD}}$ writes:
\begin{equation}
T_{\mathrm{MD}}=\frac{1}{2m\omega}(\bfk \times \bfepsilon)\cdot
(\hbar  \bfsigma + \bfL)
\end{equation}
where $\bfL=\bfr \times \bfp$ and  $\bfsigma$ is the vector of Pauli matrices. It
is proportional to the total magnetic moment operator 
$(\hbar  \bfsigma + \bfL)=( 2 \bfS+\bfL)$ where $\bfS$ is the spin operator. 
$T_{\mathrm{MD}}$ is also present in common 
non-relativistic derivations.
\cite{brouder_angular_1990,dimatteo_detection_2005}
Its selection rules are  $l_i=l_f$ and $n_i=n_f$,\cite{brouder_angular_1990} 
so it vanishes in the X-ray energy range because the states involved in the 
transition have different principal quantum numbers.

The correction to this term due to the static vector potential $\bfA_0$ is:
\begin{equation}
T_{{A_0}}=-\frac{e}{2m\omega}(\bfk \times \bfepsilon)\cdot
(\bfr\times\bfA_0).
\label{Ta0}
\end{equation}

The last term in Eq.~\eqref{TFW} is only present  when relativistic effects are included in 
the calculation of the transition operator:
\begin{equation}
T_{\mathrm{SP}}=- \frac{\hbar}{4m^2c^2} 
(\bfp-e\bfA_0)
   \cdot(\bfepsilon\times\bfsigma).
\end{equation}
A similar term was already found in Ref.~\onlinecite{GougoussisPhD}, but derived from 
a semi-relativistic Hamiltonian and this approach presents a conflict with time-dependent perturbation
theory.\cite{bouldi_dirac}
It can be rewritten noticing that, in the non-relativistic
 limit, $|\phi_i^{\mathrm{FW}}\rangle$ and $|\phi_f^{\mathrm{FW}}\rangle$
are eigenstates of:
\begin{equation}
H_0^0=\frac{(\bfp-e\bfA_0)^2}{2m}
 + eV(\bfr)- \frac{e \hbar}{2m}\mathbf{\bfsigma}\cdot\bfB_0.
\end{equation}
where $\bfB_0$ is the static external magnetic field.
This Hamiltonian obeys
$\bfp-e\bfA_0=(m/i\hbar) [\bfr,H_0^0]$ so that,
\begin{align*}
- \frac{\hbar}{4m^2c^2}
\langle \phi_f^{\mathrm{FW}}|&(\bfp-e\bfA_0)\cdot(\bfepsilon\times\bfsigma)
|\phi_i^{\mathrm{FW}}\rangle\\&=
 \frac{i}{4mc^2} (E_i-E_f)
\langle \phi_f^{\mathrm{FW}}|\bfr\cdot(\bfepsilon\times\bfsigma)|
\phi_i^{\mathrm{FW}}\rangle
\\&=
\frac{i\hbar\omega}{4mc^2}
\langle \phi_f^{\mathrm{FW}}|(\bfepsilon\times\bfr)\cdot\bfsigma|
\phi_i^{\mathrm{FW}}\rangle.
\end{align*}
We name $\bfsigma \cdot (\bfepsilon \times \bfr)$, the \textit{spin-position} 
operator and define the associated transition operator:
\begin{equation}
T_{\mathrm{SP}}=\frac{i\hbar\omega}{4mc^2}
\bfsigma \cdot (\bfepsilon \times \bfr).
\label{TSP}
\end{equation}

For technical reasons, in the present paper we consider a different situation than 
in Ref.~\onlinecite{bouldi_dirac}: we 
use FW wave function for the final states and large
components of the Dirac wave function for the initial (core)
state. This difference in treatment is linked to the fact that the core 
wave function is determined from a relativistic atomic code 
whereas the unocupied states are calculated with 
a semi-relativistic condensed-matter code.

\subsection{Rewriting the cross section with large components of the Dirac wave function for the core state}

We note $\phi_i$ and $\chi_i$ the large and small components 
of  $  \Psi_i $.
The order of magnitude of the ratio between small and large components is 
$v/c$ where $v$ is the velocity of the particle.\cite{strange_relativistic_1998}
Up to order $c^{-1}$, the small component writes:
\cite{strange_relativistic_1998,lenthe_construction_1996}
\begin{equation}
\chi_i=\frac{1}{2mc}\bfsigma \cdot (\bfp-e\bfA_0) \phi_i.
\label{chil_first}
\end{equation}

Only the second term in $U_{\mathrm{FW}}$ Eq.~\eqref{Ueriksen} couples the small
and the large components.
From Eq.~\eqref{chil_first} and \eqref{Ueriksen}, the large component of the FW 
transformed wave function can be expressed as a function of the large components of the Dirac wave function
up to order $c^{-2}$,
\begin{equation}
\phi_i^{\mathrm{FW}}=(1-\frac{1}{8m^2c^4}[\calO^2]_p)\phi_i+\frac{1}{4mc^3}
\calO_p \bfsigma \cdot (\bfp-e\bfA_0) \phi_i.
\end{equation}
$[\calO^2]_p$ is the projection of $\calO^2$ on large components:
\cite{strange_relativistic_1998}
\begin{align*}
[\calO^2]_p&=c^2(\bfp-e\bfA_0)^2-c^2e\hbar \bfsigma .\bfB_0 \\
&=2mc^2( H_0^0 - e V(\bfr))
\end{align*}
and 
$\calO_p=c \bfsigma .(\bfp-e\bfA_0)$ is the projection of $\beta \calO$ 
on the upper right components.

The identity
$c\calO_p\bfsigma \cdot (\bfp-e\bfA_0)=[\calO^2]_p$
leads to:
\begin{equation}
\phi_i^{\mathrm{FW}}=(1+\frac{1}{8m^2c^4}[\calO^2]_p)\phi_i.
\end{equation}
From this relation, the cross section of Eq.~\eqref{CSFW} can 
be adapted to the case that we consider here.  

In Ref.~\onlinecite{bouldi_dirac} the expansion was made
to order $1/c^2$ for the dipole contribution and to order $kr$
for multipole contributions.
At the same order,
\begin{align}
\sigma &= 4\pi^2\alpha_0\hbar\omega
\sum_f |\langle \phi_f^{\mathrm{FW}}|T_{\mathrm{FW}}'|\phi_i\rangle|^2
\delta(E_f-E_i-\hbar\omega)
\label{Crosssection}
\end{align}
where $T_{\mathrm{FW}}'$ is:
\begin{align*}
T_{\mathrm{FW}}' &= T_{\mathrm{FW}}(1+\frac{1}{8m^2c^4}[\calO^2]_p)\nonumber \\
&= T_{\mathrm{FW}}+T^e.
\end{align*}
There is one extra terms in the cross section compared to $T_{\mathrm{FW}}$ 
that is related to the use of large components of the Dirac wave function instead of 
Foldy-Wouthuysen wave function for the core state:
\begin{equation}
T^e=\frac{1}{2mc^2} \left( {\bfepsilon\cdot\bfr}~H_0^0- e\bfepsilon\cdot\bfr~V(\bfr) \right) .
\label{TH0}
\end{equation}

We show in the next subsection that it is  negligible 
for the core states considered in this work. 

As the magnetic dipole term is negligible in the X-ray range,
 $T_{\mathrm{FW}}$ thus contains four operators (see Eq.~\eqref{TFW}
 and the subsequent comments) so $T_{\mathrm{FW}}'$
 writes,
\begin{equation}
T_{\mathrm{FW}}'=T_{\mathrm{D}}+T_{\mathrm{Q}}+T_{{a_0}}+T_{\mathrm{SP}} +T^e.
\label{TSR}
\end{equation}

\subsection{Order of magnitude of the operators}

\begin{table*}\centering
\begin{tabular}{@{\extracolsep{4pt}}llcccccc}
\hline
\hline
\multirow{2}{*}{Edge} & &\multicolumn{1}{c}{\textit{L}$_\mathit{1}$} & 
\multicolumn{3}{c}{\textit{L}$_\mathit{2}$}& \multicolumn{2}{c}{\textit{K}} \\
\cline{3-3} \cline{4-6}\cline{7-8}
 & & I & Fe & Gd & Bi & O & Fe\\

Energy (keV) & & 5.19 & 0.72 & 7.898 & 15.71 & 0.53 & 7.11 \\[5pt]
$Z_{\rm eff}$ &  & 39.067 & 22.089 & 29.8527 & 39.2335 & 7.6579 & 25.381 \\[4pt]
$T_{\mathrm{Q}}$ \eqref{TQ}& $k r_c/2$ & $2.7{\times} 10^{-2}$ & $6.6{\times} 10^{-3}$ & $5.3{\times} 
10^{-2}$ & $8.1{\times} 10^{-2}$ &$1.4{\times} 10^{-2}$ & $5.7{\times} 10^{-2}$ \\

$T_{\mathrm{SP}}$ \eqref{TSP}& $\hbar \omega/4 mc^2$  & $2.6{\times} 10^{-3}$ &  $3.5{\times} 10^{-4}$ & $3.9 
{\times} 10^{-3}$ & $7.7{\times} 10^{-3}$ & $2.6{\times} 10^{-4}$ & $3.5 {\times} 10^{-3}$ \\

$T_{{A_0}}$ \eqref{Ta0}  & $e k r_c B_0/4m\omega$ & $6.0{\times} 10^{-6}$ & $1.1{\times} 10^{-5}$ & $7.8{\times} 
10^{-6}$ & $6.0{\times} 10^{-6}$ &$3.0{\times} 10^{-5}$ & $9.2{\times} 10^{-6}$ \\[5pt]

$T^e$ \eqref{TH0} & $(E_i- e V)/2mc^2$ & $1.3 {\times} 10^{-2}$ & $4.3 {\times} 10^{-3}$ & $7.9 {\times} 10^{-3}$ & 
$1.3 {\times} 10^{-2}$  &  $
5.2{\times} 10^{-4}$ & $5.7{\times} 10^{-3}$ \\
\hline
\hline
\end{tabular}
\caption{Order of magnitude of the operators in Eq.~\eqref{TSR} evaluated at the 
core state radius $r_c$ compared to the electric dipole operator. The mean radius of core 
orbitals is deduced from the effective nuclear charge: $r_c=\frac{3}{2}\frac{a_0}{Z_{\rm eff}}$. 
\cite{clementi_atomic_1963,clementi_atomic_1967} $B_0 $ has been fixed to $2{\times}10^4$ T (1.2eV) which is two order 
of magnitude larger than the exchange splitting observed for Fe \textit{K}-edge. The Coulomb potential is 
$V=\frac{-Z_{\rm eff}e}{4\pi \epsilon_0 r_c}$  and the core state energy $E_i$ 
is evaluated in a planetary model $E_i=\frac{-Z_{\rm eff}e^2}{8\pi \epsilon_0 r_c}$.}
\label{orderofmagnitude}
\end{table*}

As the core wave function is very localized, we obtain an idea of the 
relative order of magnitude of the operators in Eq. \eqref{TSR}  by evaluating them at the radius 
corresponding to the core state. In Table \ref{orderofmagnitude} these 
evaluations are given compared to the electric dipole operator.

When expanding the square modulus of the matrix elements in Eq.~\eqref{Crosssection}, we keep the terms with 
contributions higher than $10^{-3}$ compared to the dominant electric dipole 
term. We also neglect the term $T^e$: 
as $V(\bfr)$ is almost spherical at the core state radius, 
it concerns transitions to the same 
orbitals as the electric dipole term. It does not contain a spin operator 
so that, even in XMCD, it only yields a negligible correction to the 
electric dipole contribution. Therefore, we are left with the four significant terms D-D, Q-Q, D-Q and D-SP discussed in section~\ref{sec:contributions}.

\end{document}